\documentstyle[preprint,aps,epsf,psfig]{revtex}
\begin{document}
\draft
\preprint{\begin{tabular}{l}
\hbox to\hsize{October, 1998 \hfill KAIST-013/98}\\[-3mm]
\hbox to\hsize{hep-ph/9810366 \hfill SNUTP 98-077}\\[5mm] \end{tabular} }
\vspace{1cm}
\title{B decays into two light pseudoscalar mesons \\
and possible effects of enhanced $b\rightarrow s g$ 
}
 
\vspace{2cm}
   
\author{ 
Ji-Ho Jang, \footnote{e-mail~:~jhjang@chep6.kaist.ac.kr} 
Yeong Gyun Kim  \footnote{e-mail~:~ygkim@chep6.kaist.ac.kr} 
and Pyungwon Ko  \footnote{e-mail~:~pko@charm.kaist.ac.kr} }
\address{
Department of Physics, Korea Advanced Institute of Science and Technology,\\
Taejon 305-701, Korea}
\maketitle
            
\begin{abstract}
We consider the branching ratios and $CP$ asymmetry in the $B$ meson decays 
into two pseudoscalar mesons in the generalized factorization approximation. 
We also investigate the possible effects of the enhanced chromomagnetic 
interaction $b\rightarrow s g$ on these exclusive $B$ meson decays that was
suggested as a possible solution to the semileptonic branching ratio of $B$
mesons, the missing charm puzzle and the large $B \rightarrow \eta^{'} + X_s$.
We finds that such enhanced $b\rightarrow s g$ interaction degrades the 
agreement between the data and the model predictions for $B$ meson decays into
two light mesons in the generalized factorization approximation. 
\end{abstract}


\newpage

\narrowtext
\tighten

\section{Introduction}
Recently, the CLEO collaboration reported the observation of some 
nonleptonic $B$ meson decays into two light pseudoscalar mesons 
with the following branching ratios  \cite{CLEOpp,CLEOeta} :
\begin{eqnarray}
{\cal B} ( B^{\pm} \rightarrow \pi^{\pm} K^0 ) &=&
 ( 2.3^{+1.1}_{-1.0} \pm 0.3 \pm 0.2 ) \times 10^{-5} \nonumber \\
{\cal B} ( B_d \rightarrow \pi^{\pm} K^{\mp} ) &=&
 ( 1.5^{+0.5}_{-0.4} \pm 0.1 \pm 0.1 ) \times 10^{-5} \nonumber \\
{\cal B} ( B^{\pm} \rightarrow K^{\pm} \eta' ) &=&
 ( 6.5^{+1.5}_{-1.4} \pm 0.9 ) \times 10^{-5}  \nonumber \\
{\cal B} ( B_d \rightarrow K^0 \eta' ) &=&
 ( 4.7^{+2.7}_{-2.0} \pm 0.9 ) \times 10^{-5}, 
\end{eqnarray}
And the upper limits of the branching ratios for other decay modes were of 
the order of $10^{-5}$. 
These measurements at CLEO and the future experiments at B-factories 
motivate a great deal of theoretical interests in nonleptonic $B$ decays  
into two light mesons \cite{DG97,AG97,ACGK97,DHP97,CT97,DGR97,DDO97,ali98}.

The standard theoretical framework to study non-leptonic $B$ decays
is based on the effective Hamiltonian approach.
The short and long distance QCD effects in the non-leptonic weak decays
are separated by means of the operator product expansion\cite{Wil69}.
The resulting
effective Hamiltonian consists of products of scale-dependent Wilson
coefficients $C_i ( \mu )$ and the local four-quark operators at the 
renormalization scale $\mu$. The short distance contributions of the
coefficients $C_i ( \mu )$ has been evaluated up to the next-to-leading 
logarithmic ( NLL ) order \cite{CMM97,BJLW92}. 
In the NLL precision, the Wilson coefficients  
depend on the renormalization scheme as well as renormalization scale. 
These dependences should be canceled by corresponding scheme/scale dependence
of the matrix element of the operators. 
However, hadronic matrix elements are usually calculated under the 
factorization approximation in which they are replaced by the scale/scheme 
independent form factors and decay constants.
In order to achieve the scale/scheme independence of the matrix elements
of the effective Hamiltonian,
Ali and Greub \cite{AG97} included the corrections of 
the one-loop penguin-like diagrams and
some process independent part of the vertex corrections associated
with the four-fermi operators to the partonic matrix elements 
before doing factorization step.
They also have taken into account the effects of the $O(\alpha_s)$ 
tree-level matrix element associated with the chromomagnetic dipole operator.
Such corrections compensate the scale/scheme dependences of the Wilson
coefficients and one finally get the effective Hamiltonian involving
the scale/scheme independent Wilson coefficients and four-quark operator.

The factorization method is successful to describe the heavy-to-heavy
nonleptonic decays \cite{neubert}. The intuitive argument for the
factorization is given by Bjorken \cite{bjorken} based on the 
idea of color transparency. The quark pairs with high energy coming from 
the heavy meson decay hadronize after they have traveled some distance
from each other. Hence the decay process is expected to factorize
into the color singlet current pairs since soft gluon effects are small 
after the hadronization occurs.
There is  more theoretical argument about the factorization based on QCD.
In the specific kinematic region in which the two light 
quarks are highly collinear and all quarks are almost on-shell, 
the leading term of the Green function in expansion of
inverse powers of the heavy quark masses and the large energy transferred
to the light quark pairs exhibits factorization  \cite{dg91}.
The authors in Ref.~ \cite{AG97} extended this framework to 
the heavy-to-light transitions.
They also introduce a new free parameter $\xi$ that describes 
nonfactorization effects such as color octet contributions and concluded 
that the range $0 \leq \xi \leq 0.5$ is consistent with data \cite{AG97}.
However one should keep in mind that the factorization for the 
heavy-to-light transition is still only a model to describe the complex 
hadronic matrix elements since there is no theoretical justification
as the heavy-to-heavy transition case \cite{dg91}.
In principle there is no reason that only one single value of the parameter
$\xi$ can explain the branching ratios of all kind of different modes.
For example, the authors of Ref.~\cite{CT97} introduce two different $\xi$'s 
corresponding to the different currents structures which give different
nonfactorization corrections. 
However, we will use the factorization method including one 
nonfactorization parameter $\xi$ in this work.

On the other hand, there are some possible anomalies in the inclusive
$B$ decays. The persistent discrepancy of the measurements of the 
semileptonic branching ratio and charm multiplicity with the theoretical 
predictions have been known quite for a while \cite{bpuzzle}.  One can argue
that these problems arise because of the breakdown of local quark-hadron 
duality that was invoked when one estimates the nonleptonic $B$ decay rate. 
However, several authors have noticed that these puzzles can 
be solved if one assumes that the Wilson coefficient of the chromomagnetic 
operator is enhanced by some new physics contributions \cite{hou}.
Incidentally, this may help to understand the recently  measured branching 
ratio for ${\cal B}(B \rightarrow \eta^\prime X_s)$, which is surprisingly 
larger than previously expected.  Such an enhanced $b\rightarrow s g$ should 
affect inevitably the exclusive decay rates of $B$ mesons, and it is one
of the main themes of our present work to study such effects. 

In this work, we consider the nonleptonic $B$ decays into two light 
pseudoscalar mesons and investigate the possible effects through
the enhanced chromomagnetic dipole contribution. The branching ratios are 
calculated by the generalized factorization method proposed in Ref.~ 
\cite{AG97}. The QCD and EW penguin contributions are included in this work. 
The two-angle mixing formalisms are used in the calculation of the decay 
rates involving $\eta, \eta'$ mesons, and the amplitudes
$b \rightarrow s ( gg ) \rightarrow s ( \eta, \eta' )$ \cite{ACGK97} are
included using the QCD anomaly, instead of the intrinsic charm contents of
$\eta^{(')}$. We also study the CP asymmetry in $B$ meson decays,
both the direct $CP$ asymmetries of charged $B$ meson decays and 
the time integrated $CP$ asymmetries of neutral $B$ meson decays.
In this calculation, we assume that the 
strong phases are given by the penguin-type diagrams with internal light 
quarks which could have on-shell momentum 
\cite{BSS79}.

\section{Calculational Framework}
The $\Delta B = 1$ effective Hamiltonian is given by
\begin{equation}
H_{eff}= \frac{G_F}{\sqrt{2}} 
\sum_{q=d,s} \left[ V_{ub} V_{uq}^* ( C_1 O^u_1 + C_2 O^u_2 ) +
 V_{cb} V_{cq}^* ( C_1 O^c_1 + C_2 O^c_2 ) -
 V_{tb} V_{tq}^* \left( \sum_{i=3}^{10} C_i O_i + C_g O_g \right)
 \right]
\end{equation}
where the operators are
\begin{equation}
\begin{array}{l}
\begin{array}{ll}
O^u_1= ( \bar{u}_{\alpha} b_{\alpha} )_{V-A} 
         ( \bar{q}_{\beta} u_{\beta} )_{V-A}&
O^c_1= ( \bar{c}_{\alpha} b_{\alpha} )_{V-A} 
         ( \bar{q}_{\beta} c_{\beta} )_{V-A}\\
O^u_2= ( \bar{u}_{\beta} b_{\alpha} )_{V-A} 
         ( \bar{q}_{\alpha} u_{\beta} )_{V-A}&
O^c_2= ( \bar{c}_{\beta} b_{\alpha} )_{V-A} 
         ( \bar{q}_{\alpha} c_{\beta} )_{V-A}\\
O_3= (\bar{q}_{\alpha} b_{\alpha} )_{V-A}
      \sum_{q'}\left( \bar{q}^{'}_{\beta} q^{'}_{\beta} \right)_{V-A}&
O_4= (\bar{q}_{\beta} b_{\alpha} )_{V-A}
      \sum_{q'}\left( \bar{q}^{'}_{\alpha} q^{'}_{\beta} \right)_{V-A}\\
O_5= (\bar{q}_{\alpha} b_{\alpha} )_{V-A}
      \sum_{q'}\left( \bar{q}^{'}_{\beta} q^{'}_{\beta} \right)_{V+A}&
O_6= (\bar{q}_{\beta} b_{\alpha} )_{V-A}
      \sum_{q'}\left( \bar{q}^{'}_{\alpha} q^{'}_{\beta} \right)_{V+A}\\
O_7= \frac{3}{2} (\bar{q}_{\alpha} b_{\alpha} )_{V-A}
      \sum_{q'} e_{q'} \left( \bar{q}^{'}_{\beta} q^{'}_{\beta} \right)_{V+A}&
O_8=\frac{3}{2}  (\bar{q}_{\beta} b_{\alpha} )_{V-A}
   \sum_{q'} e_{q'} \left( \bar{q}^{'}_{\alpha} q^{'}_{\beta} \right)_{V+A}\\
O_9= \frac{3}{2} (\bar{q}_{\alpha} b_{\alpha} )_{V-A}
      \sum_{q'} e_{q'} \left( \bar{q}^{'}_{\beta} q^{'}_{\beta} \right)_{V-A}&
O_{10}=\frac{3}{2}  (\bar{q}_{\beta} b_{\alpha} )_{V-A}
      \sum_{q'} e_{q'} \left( \bar{q}^{'}_{\alpha} q^{'}_{\beta} \right)_{V-A}
\end{array} \\
O_g=(g_s/8 \pi^2)~ m_b~ \bar{s}_{\alpha} \sigma^{\mu\nu}
     ( 1+ \gamma_5 )~( \lambda^A_{\alpha\beta}/2 ) b_{\beta}~ G^A_{\mu\nu}.
\end{array}
\end{equation}
Here the $O_1$ and $O_2$ are the current-current operators. 
The $O_3 \sim O_6$ and $O_7 \sim O_{10}$ are called the gluonic and 
electroweak penguin operators respectively, and $O_g$ is the gluonic dipole 
moment operator. The subscripts $V \pm A$ represent the projection 
operators $1 \pm \gamma_5$ onto left- and right-handed spinor and $e_{q'}$
indicates the electromagnetic charge of the corresponding quarks.
The subindex $\alpha$ and $\beta$ are the $SU(3)$ color indices and 
$\lambda^A_{\alpha\beta}~( A=1 \sim 8)$ are the Gell-Mann matrices.

If we take $m_{top}^{pole} = 175$ GeV, $\alpha_s ( M_Z ) = 0.118, \alpha 
( M_Z ) = 1/128$, we have the following numerical values of the Wilson 
coefficients  at the renormalization scale $\mu = 2.5$ GeV in the 
naive dimensional renormalization scheme \cite{AG97}:
\begin{equation}
\begin{array}{llll}
C_1 = 1.117 & C_2 = -0.257 & C_3 = 0.017 & C_4 = -0.044 \\
C_5 = 0.011 & C_6 = -0.056 & C_7 = -1 \times 10^{-5} &C_8 = 5 \times 10^{-4} \\
C_9 = -0.010 & C_{10} = 0.002 & C_g = -0.158 &
\end{array}
\end{equation}
where the Wilson coefficients $C_1 \sim C_6$ are taken as the
NLL values with respect to QCD. The values
of the remaining coefficients are given at the leading logarithmic
precision. 

In the NLL precision, the matrix elements of local 4-fermion operators 
$O_i$'s are to be treated at the one-loop level.
In Ref.\cite{AG97}, the contributions arising from the penguin-type
diagram of the operators $O_1 \sim O_6$ and the tree-level diagram
of the dipole operator $O_g$ have been calculated and absorbed into
the effective Wilson coefficients $C_i^{eff}$.
The process-independent contributions from the vertex-type diagrams are 
also considered. These full NLL considerations are sufficient to 
compensate the scale and scheme dependence arising from the
factorization approximation and the resulting 
effective Wilson coefficients are given in Ref.\cite{AG97,ACGK97}.

In this paper, we use the factorization approximation in order to
calculate the hadronic matrix elements of the type $<h_1 h_2|O_i|B>$,
where $h_{1,2}$ is a light pseudoscalar meson.
In this approximation, the hadronic matrix elements are factorized into
a product of two matrix elements of quark bilinear operators. 
As an example, let us consider the decay ${\bar B}^0 \rightarrow \pi^+ 
\pi^-$, whose matrix element is given by

\begin{eqnarray}
M&=& \frac{G_F}{\sqrt{2}} \left\{ V_{ub} V_{ud}^*  a_1
- V_{tb} V_{td}^* \left[ a_4 + a_{10}
+\frac{ 2 m_{\pi}^2 ( a_6 + a_8 )}{( m_u + m_d )( m_b - m_u )}
\right] \right\} \nonumber \\
&& \times
< \pi^{-} | \bar{d} u_{-} | 0 >< \pi^+ | \bar{u} b_{-} | \bar{B}^0 >
\end{eqnarray}
where
\begin{eqnarray}
a_i = C_i^{eff} + \xi~ C_{i+1}^{eff} ( i = odd )~~~,
a_i = C_{i+1}^{eff} + \xi~ C_i^{eff} ( i = even ).
\end{eqnarray}
Here, instead of $1/N_c$, we introduce a free parameter $\xi$ which is
supposed to describe the nonfactorization contribution. 
(The amplitudes for  other decays are explicitly listed in Appendix.)

When we express the matrix elements of quark bilinear operators in terms 
of meson decay constants and form factors, it is important to remember 
which convention we use for the meson wave function in terms of quark flavor
contents. We adopt the following convention, for which 
the meson state is the same as the isospin eigenstate without extra sign. 

\begin{eqnarray}
&|&\pi^+> = |u \bar{d}>,~~
|\pi^0> = |\frac{1}{\sqrt{2}} (d \bar{d} - u \bar{u})>,~~
|\pi^-> = |u \bar{d}> \nonumber \\
&|&K^-> = |-s \bar{u}>,~~
|{\bar K}^0> = |s \bar{d}>,~~ |K^0> = |d \bar{s}>,
~~ |B^-> = |-b \bar{u}>,~~
|{\bar B}^0> = |b \bar{d}> \nonumber \\
&|&\eta_8> = |\frac{1}{\sqrt{6}} (2 s\bar{s} - u \bar{u} - d\bar{d})>,~~
|\eta_0> = |\frac{1}{\sqrt{3}} (-s\bar{s} - u\bar{u} - d\bar{d})> 
\end{eqnarray}
With this convention, we define the decay constants and form factors
as follows:
\begin{eqnarray}
&<&\pi^-(p)|\bar{d}u_-|0>=i f_{\pi} p_{\mu},~~~
<K^-(p)|\bar{s}u_-|0>=i f_{K} p_{\mu} \nonumber \\
&<&\eta_8(p)|\frac{1}{\sqrt{6}}(\bar{u}u_-+\bar{d}d_- -2\bar{s}s_-)|0>
=i f_8 p_\mu \nonumber \\
&<&\eta_0(p)|\frac{1}{\sqrt{3}}(\bar{u}u_-+\bar{d}d_- +\bar{s}s_-)|0>
=i f_0 p_\mu \nonumber \\
&<&\eta^{(\prime)}(p)|\bar{u}u_-|0>=if^u_{\eta^{(\prime)}} p_\mu,~
<\eta^{(\prime)}(p)|\bar{s}s_-|0>=if^s_{\eta^{(\prime)}} p_\mu,~
<\eta^{(\prime)}(p)|\bar{c}c_-|0>=if^c_{\eta^{(\prime)}} p_\mu 
\end{eqnarray}
and
\begin{eqnarray}
&<&\pi^-(p^\prime)|\bar{d} b_-|B^-(p)>=
[ (p+p^\prime)_\mu - \frac{m_B^2-m_\pi^2}{q^2} q_\mu ]
F_1^{B \rightarrow \pi} (q^2) +
\frac{m_B^2-m_\pi^2}{q^2} q_\mu F_0^{B \rightarrow \pi} (q^2) \nonumber \\
&<&K^-(p^\prime)|\bar{s}b_-|B^-(p)>=
[ (p+p^\prime)_\mu - \frac{m_B^2-m_K^2}{q^2} q_\mu ]
F_1^{B \rightarrow K} (q^2) +
\frac{m_B^2-m_K^2}{q^2} q_\mu F_0^{B \rightarrow K} (q^2) \nonumber \\
&<&\eta^{(\prime)}(p^\prime)|\bar{u} b_-|B^-(p)>=
[ (p+p^\prime)_\mu - \frac{m_B^2-m_{\eta^{(\prime)}}^2}{q^2} q_\mu ]
F_1^{B \rightarrow {\eta^{(\prime)}}} (q^2) +
\frac{m_B^2-m_{\eta^{(\prime)}}^2}{q^2} 
q_\mu F_0^{B \rightarrow {\eta^{(\prime)}}} (q^2) 
\end{eqnarray}
Other matrix elements are related to the above matrix elements
using the following isopin relations.
\begin{eqnarray}
&<&\pi^-(p)|\bar{d}u_-|0>=
\sqrt{2} <\pi^0|\bar{u}u_-|0>=
-\sqrt{2} <\pi^0|\bar{d}d_-|0> \nonumber \\
&<&K^-(p)|\bar{s}u_-|0>=-<{\bar K}^0|\bar{s}d_-|0> =-<K^0|\bar{d}s_-|0>,~~
<\eta^{(\prime)} | \bar{u}u_- | 0 > = 
<\eta^{(\prime)} | \bar{d}d_- | 0 > \nonumber \\
&<&\pi^-(p^\prime)|\bar{d} b_-|B^-(p_B)>=<\pi^+|\bar{u} b_-|{\bar B}^0>=
\sqrt{2} <\pi^0|\bar{d} b_-|{\bar B}^0>=
\sqrt{2} <\pi^0|\bar{u} b_-|B^-> \nonumber \\
&<&K^-(p^\prime)|\bar{s}b_-|B^-(p_B)>
=<{\bar K}^0|\bar{s}b_-|{\bar B}^0>,~~
<\eta^{(\prime)} | \bar{u} b_- | B^- >
= -<\eta^{(\prime)} | \bar{d}b_- |{\bar B}^0> 
\end{eqnarray}

We choose the nemerical values of the relevant decay constants,
$f_{\pi}, f_K$ as follows:
\begin{equation}
f_{\pi}=131~MeV,~~f_K= 160~MeV.
\end{equation} 
The values of form factors at $q^2=m_h^2$ is required for
the calculation of decay rate. As the final states involve 
only light hadrons we can safely neglect the $q^2$ dependence of
form factors. Hence we assume that $F_{0,1}( m_h^2 ) = F_{0,1}( 0 )$.
The values used for the form factors are obtained using the BSW model
\cite{BSW87},
\begin{equation}
\begin{array}{ll}
F_{0,1}^{B \rightarrow K } = 0.33,~~ & {\rm and}~~
F_{0,1}^{B \rightarrow \pi} = 0.33 
\end{array}
\end{equation}

In the case with nonleptonic $B$ decays into the final state involving 
$\eta$ or $\eta^{'}$, we use two-angle mixing formalism developed by 
Leutwyler \cite{Leut97}. This formalism is phenomenologically adequate to 
explain the various experimental data, the decay width $\Gamma ( \eta 
\rightarrow 2 \gamma )$, $\Gamma ( \eta' \rightarrow 2 \gamma )$  and the 
ratio $\Gamma ( J/\psi \rightarrow \eta' \gamma )/\Gamma ( J/\psi 
\rightarrow \eta \gamma )$, as recently discussed in Refs.~\cite{Leut97,FK97}. 
In this formalism, the $\eta$ and $\eta'$ states are defined by the mixture of 
the octet and singlet states with different mixing angles such as 
\begin{equation}
|\eta > = \cos \theta_8 |\eta_8 > - \sin \theta_0 | \eta_0 >, ~~~~
|\eta' > = \sin \theta_8 |\eta_8 > + \cos \theta_0 | \eta_0 >.
\end{equation}
Then, the decay constants $f^{u}_{\eta^{(')}}, f^{s}_{\eta^{(')}}$ 
are obtained through the
two mixing angles and $f_{8}, f_{0}$:
\begin{equation}
\label{decayconstant}
\begin{array}{ll}
f^u_{\eta}=\frac{f_8}{\sqrt{6}} \cos \theta_8 - 
           \frac{f_0}{\sqrt{3}} \sin \theta_0, &
f^u_{\eta'}=\frac{f_8}{\sqrt{6}} \sin \theta_8 + 
           \frac{f_0}{\sqrt{3}} \cos \theta_0.  \\
f^s_{\eta}=-2 \frac{f_8}{\sqrt{6}} \cos \theta_8 - 
           \frac{f_0}{\sqrt{3}} \sin \theta_0, &
f^s_{\eta'}=-2 \frac{f_8}{\sqrt{6}} \sin \theta_8 + 
           \frac{f_0}{\sqrt{3}} \cos \theta_0.
\end{array}
\end{equation}
And the relevant form factors for $B \rightarrow \eta$ and
$B \rightarrow \eta^{\prime}$ are:
\begin{equation}
\begin{array}{ll}
F_{0,1}^{B \rightarrow \eta} = F_{0,1}^{B \rightarrow \pi} 
\left[\frac{\cos \theta_8}{\sqrt{6}} - \frac{\sin \theta_0}{\sqrt{3}} \right] 
,&
F_{0,1}^{B \rightarrow \eta'} = F_{0,1}^{B \rightarrow \pi} 
\left[\frac{\sin \theta_8}{\sqrt{6}} + \frac{\cos \theta_0}{\sqrt{3}} \right]
.
\end{array}
\end{equation}
For numerical analysis, we choose the following values:
\begin{equation}
\theta_8=-22.2^o,~~\theta_0=-9.1^o,
~~\frac{f_8}{f_\pi}=1.28,~~\frac{f_0}{f_\pi}=1.20
\end{equation}
which are given by fitting of data on
the decay width $\Gamma ( \eta \rightarrow 2 \gamma )$, 
$\Gamma ( \eta' \rightarrow 2 \gamma )$  and the ratio 
$\Gamma ( J/\psi \rightarrow \eta' \gamma )
/\Gamma ( J/\psi \rightarrow \eta \gamma )$ \cite{FK97}. The parameters 
$f_{\eta}^{( c )}, f_{\eta'}^{( c )}$ quantify  the contributions from the 
decay $b \rightarrow s ( c\bar{c} ) \rightarrow s ( \eta, \eta' )$. Their 
magnitude and the sign were estimated using the QCD anomaly method in Ref. 
\cite{ACGK97}, and we use their values in this article : $f_{\eta}^c = -0.9$ 
MeV and $f_{\eta'}^c = -2.3$ MeV for $m_c = 1.5$ GeV. 

For numerical calculation of the transition matrix elements,
we also need numerical values of CKM elements and quark masses.
We use the Wolfenstein parameterization \cite{Wol83}
of CKM matrix element.
Two parameters $A, \lambda$ are well determined using 
$|V_{cb}|$ through the fitting of the $B \rightarrow D^{*} l \nu_l$ decay
spectrum and $|V_{us}|$ through the 
$K \rightarrow \pi e \nu$ and hyperon decays :
$A=0.81 \pm 0.06$ and $\lambda=\sin \theta_c = 0.0025 \pm 0.0018$.
In this paper, we choose the central value of these parameters.
Other two parameters in CKM matrix elements
are constrained by CKM unitarity fitting \cite{Ali96} as
$0.025 \leq \eta \leq 0.52$ and $-0.25 \leq \rho \leq 0.35$ 
( 95 \% C.L. ). However the lower bound on the mass mixing ratio
$\Delta M_s$ / $\Delta M_d$ and the experimental value of 
$R_1 \equiv  B ( B^0(\bar{B}^0) \rightarrow \pi^{\pm} K^{\mp})/
B ( B^{\pm} \rightarrow  \pi^{\pm} K^0 )=0.65 \pm 0.40$ disfavor
negative $\rho$ region. Using the relation, 
$\sqrt{\rho^2+\eta^2}=|V_{ub}|/\lambda |V_{cb}|$, we choose the
three typical values of $\rho$ and $\eta$ satisfying the 
above constraint following ref. \cite{ACGK97},
\begin{eqnarray}
(\rho=0.05,~ \eta=0.36),~~~ 
(\rho=0.30,~ \eta=0.42),~~~ 
(\rho=0.00,~ \eta=0.22). \nonumber
\end{eqnarray}
These correspond  $|V_{ub}| / |V_{cb}| = 0.08, 0.11$, and $0.05$
representing the center value and upper and lower limit of the 
experimental value ( 90 \% C.L. ) of the ratio.

In the calculation of Wilson coefficients, we use the internal quark
masses as constituent quark masses: $m_b = 4.88$ GeV,  $m_c = 1.5$ GeV,
$m_s = 0.5$ GeV, $m_d = m_u = 0.2$ GeV. The mass terms in the matrix
elements of several decay modes come form the equation of motion.
Hence they are current quark masses rather than the constituent 
quark masses and 
the values of them are given by $m_b = 4.88$ GeV, $m_c = 1.3$ GeV,
$m_s = 122$ MeV, $m_d = 7.6$ MeV, $m_u = 4.2$ MeV  at the renormalization 
scale $\mu = 2.5$ GeV.

\section{Branching Ratios}

Armed with the strategies described in the previous section, it is 
straightforward to calculate the amplitudes and the branching ratios of
$B$ meson decays into two light pseudoscalar mesons. The full amplitudes are
given in the Appendix, and we will consider the numerical results only in 
this section. We present the results for the averaged branching ratio, since
CLEO has measured the averaged one : for example,
\begin{equation}
{\cal B} ( B^{\pm} \rightarrow \eta \pi^{\pm} )
 \equiv \frac{1}{2} \left[
{\cal B} ( B^- \rightarrow \eta \pi^- )
+ {\cal B} ( B^+ \rightarrow \eta \pi^+ ) \right].
\end{equation}

In Table 1, we present the numerical results using the typical parameters
: $\xi=1/3$, $f_{\eta}^{(c)}= -0.9 $ MeV, $f_{\eta'}^{(c)}= -2.3$ MeV
( $m_c = 1.5 $GeV ), $\rho=0.05$, $\eta=0.36$. In Figs.~ 1, 2 and 3, 
we plot the branching ratio as a function of nonfactorization
parameter $\xi$ and using three different set of CKM angle $\rho, \eta$.

The additional effects of EW penguin diagrams 
in $B \rightarrow \pi \pi, \pi K, K K$ modes,
except $B_d \rightarrow \pi^0 \pi^0, \pi^0 K^0$ 
and $B^{\pm} \rightarrow \pi^0 K^{\pm}$ modes,
are generally negligible
and our predictions on the decay rates including EW penguin
on the several modes are similar to that of Ref. \cite{AG97}.
and consistent on the recent CLEO data \cite{CLEOpp}. 
In  $B^{\pm} \rightarrow K^{\pm} \eta'$, the theoretical prediction
has about $2\sigma$ deviation from the CLEO data.
In $B_d \rightarrow \pi^0 \pi^0, \pi^0 K^0$ modes,
the EW penguin effects decrease
the decay rate about $33\%$ and  $25\%$ for $\xi = 1/3$,
respectively.
In $B^{\pm} \rightarrow \pi^0 K^{\pm}$ mode,
the decay rate is increased by the effects of EW penguin diagrams
about $29\%$ for $\xi = 1/3$.
The decay rate of $B^{\pm} \rightarrow K^0 K^{\pm}$ is same as
that of $B_d \rightarrow K^0 \bar{K}^0$ 
because of the isospin symmetry. 

In $B \rightarrow \pi \eta^{(')}, K \eta^{(')}, \eta^{(')} \eta^{(')}$
modes except $B^{\pm} \rightarrow K^{\pm} \eta, 
B_d \rightarrow K^0 \eta, \eta \eta$ modes, 
the EW penguin effects are also negligible and give
at most ${\cal O} ( {\rm few}~ \% )$ corrections in decay rates.
For $B_d \rightarrow \eta \eta$ modes, the effects increase the decay
rate about $20\%$. In the $B^{\pm} \rightarrow K^{\pm} \eta, 
B_d \rightarrow K^0 \eta$, the decay rates are decreased by the effects
about $39\%$ and $36 \%$ respectively.
In $B_d \rightarrow \pi^0 \pi^0, \eta^{(')} \eta^{(')}$ modes, 
the branching ratio plot has minimum value between 
$\xi=0.2$ and $\xi=0.3$ and its values become very small in this region
and very sensitive to the actual values of $\xi$, so that we cannot trust 
our predictions too much. 

We also compare the decay rates with and without the 
$b \rightarrow s(c\bar{c}) \rightarrow s(\eta, \eta')$ contributions.
These corrections does not change the decay rates significantly.
Our results are quite different from  those obtained in Ref. \cite{HZ97},
where $b \rightarrow s(c\bar{c}) \rightarrow s(\eta, \eta')$ effect
was interpreted as the intrinsic charm contents in $\eta, \eta'$ mesons.
The authors of Ref.~\cite{HZ97} estimated the numerical value of 
$f^{(c)}_{\eta}/f^u_{\eta}$ to be ${\cal O}(1)$, and predicted that
the branching ratios are about 
$0.194 \times 10^{-5} ( 5.83 \times 10^{-5})$ for the decay modes 
$B^{\pm} \rightarrow K^{\pm} \eta ( \eta' )$, and 
$0.027 \times 10^{-5} ( 5.73 \times 10^{-5} )$ for the decays modes 
$B_d \rightarrow K^0 \eta ( \eta' )$, which are substantially larger than 
our predictions.  However, such large $f_{\eta^{(')}}^{(c)}$ are not 
compatible with other theoretical/experimental considerations 
\cite{ACGK97,FK97}, and cannot be taken too seriously. 


\section{$CP$ asymmetries}

We also consider the direct $CP$ violating rate asymmetry in the charged
$B$ meson decay and the time integrated $CP$ asymmetry in neutral
$B$ meson decay. In the inclusive charmless $B$ decays,
the necessary strong phase differences are obtained from
the absorptive part of the penguin diagram \cite{BSS79,Wol91,GH91}. 
Such method is applied to exclusive nonleptonic decays of 
$B$ meson decays into two light pseudoscalar mesons \cite{KPS95}.
It had been usually assumed that the final state interactions in nonleptonic 
$B$ decays may be negligibly small, 
because the mass of $B$ meson is far above the usual resonance region.
However such effects become important in some cases. For example,
the soft rescattering effects\cite{BFM97,Neu97,Flei98}
might affect the method to constrain and determine the CKM angle $\gamma$ 
using $B \rightarrow \pi K$ modes \cite{FM98}.

In this article, we neglect the final state interactions and 
calculate the typical values and various parameter dependences of
$CP$-asymmetries.

The $CP$-violating rate asymmetry is defined as
\begin{equation}
a_{CP}=\frac{\Gamma(B^- \rightarrow f) - \Gamma(B^+ \rightarrow \bar{f})}
            {\Gamma(B^- \rightarrow f) + \Gamma(B^+ \rightarrow \bar{f})}
\end{equation}
where  $\bar{f}$ is the chrege conjugated state of the final state $f$.

The time integrated $CP$-asymmetry of neutral $B_d$ meson \cite{Gro89} is
\begin{equation}
a_{CP}( B_d \rightarrow f ) =
\frac{1}{1+x_d^2} \left[ A_{CP}^{dir} ( B_d \rightarrow f )
 + A_{CP}^{mix-ind} ( B_d \rightarrow f ) \right],
\end{equation}
with
\begin{equation}
A_{cp}^{dir}( B_d \rightarrow f ) = \frac{1 -|\xi_f|^2}{1+|\xi_f|^2},
~~~~
A_{cp}^{mix-ind}( B_d \rightarrow f ) = \frac{- 2 x_d Im \xi_f}{1+|\xi_f|^2}.
\end{equation}
and
\begin{equation}
x_d = \frac{\Delta m_{B_d}}{\Gamma_{B_d}} \sim 0.71, ~~~~~ 
\xi_f= \frac{q}{p}~ \frac{A(\bar{B^0} \rightarrow f)}{A(B^0 \rightarrow f)}.
\end{equation}
where $f$ is $CP$ eigenstates and $\Delta m$ denotes the mass difference
of $B^0$ and $\bar{B^0}$ mesons. The ratio $q/p$ represents the CKM matrix
elements ratio contributing to the $B^0 - \bar{B^0}$ mixing.
If we consider $B$ meson decay into the final states with $K^0 ( \bar{K}^0 )$
meson, the parameter $\xi_f$ has additional factor $( q/p )_K$ for the
$K^0 - \bar{K}^0$ mixing effects.

We present the typical values of $CP$ asymmetry of several modes
and estimate the EW penguin effects and  $f_{\eta}^{(c)}$ effects
in Table 2.
We also consider the effects of different CKM angles $\rho, \eta$.
In the Figures 4,5 and 6, the $CP$ asymmetry is given by the function
of nonfactorization parameter $\xi$ with three different CKM angles
$\rho, \eta$.
In Ref \cite{KPS95}, the authors present the results of $CP$ asymmetry
as a function of $q^2$.
We will fix this value as $m_b^2/2$ in this analysis.
The EW penguin effects and 
the $b \rightarrow d ( s ) [ c \bar{c} \rightarrow g g
\rightarrow \eta^{(')}]$ type corrections on $CP$ violation
are generally small.
The two-mixing scheme also gives generally small 
corrections to the values of the
$CP$ asymmetry compared with
the one-mixing formalism for $\eta - \eta'$ system.

In the  $CP$ asymmetry in charged $B$ decays, The nonfactorization
parameter $\xi$ dependence is generally mild.
The $CP$ asymmetry in $B^{\pm} \rightarrow \pi^{\pm} \pi^0,
K^{\pm} \pi^0$ is very small, at most $ 0.2 \%$ in magnitude. 
For $B^{\pm} \rightarrow K^{\pm} \pi^0, K^{\pm} \eta$ mode, the magnitude
of $CP$ asymmetry is less than about $10 \%$ in the most parameter sets.
In $B^{\pm} \rightarrow K^{\pm} K^0, \pi^{\pm} \eta^{(')}$, 
and $ K^{\pm} \eta$ modes,
the range of the magnitude of $CP$ asymmetry is $ 10 \sim 30 \%$.
%
In the $CP$ asymmetry in neutral $B$ meson decays, 
$B_d \rightarrow \pi^0 \pi^0$ and $B_d \rightarrow \eta^{(')} \eta^{(')}$ 
modes give very steep $\xi$ dependence in
$0.2 \sim 0.4$ region of $\xi$ region. The sign of $CP$ asymmetry is 
changed in the region near $\xi \sim 0.3$. Hence the typical values 
for such modes in table 2 ( which values are given in $\xi = 1/3$ ) should
be largely changed in small shift of $\xi$.
In other mode, the $CP$ asymmetry change slowly in varying $\xi$.


\section{Possible effects through enhancements in $b\rightarrow s g$} 

For the last several years, there have been some speculations about the 
enhanced $b \rightarrow sg$ by several authors \cite{hou} in order to 
resolve some discrepancies in the data and theoretical expectations in 
inclusive $B$ decays. 
The CLEO and ARGUS collaboration \cite{BHP96,Barish96} 
have measured the semileptonic branching ratio:
\begin{equation}
\label{bslexp}
B_{SL}^{exp}=10.23 \pm 0.39 \%
\end{equation}
And the CLEO 1.5, CLEO II and ARGUS data\cite{BHP96} give
\begin{equation}
\label{ncexp}
n_c^{exp}=1.15 \pm 0.05
\end{equation}
for the average number of charm(anti-) quarks per $B^+/B^0$-decay.
The result (\ref{bslexp}) is considerably smaller than 
the theoretical prediction 
in the parton model. It has been found that charm mass corrections to
$\Gamma(b \rightarrow c\bar{c}s)$ are large and can reduce the theoretical
prediction for $B_{SL}$ \cite{BBBG94}:
\begin{equation}
\label{bslth}
B_{SL}=(11.7 \pm 1.4 \pm 1.0)\%
\end{equation}
Then, there is no spectacular discrepancy between (\ref{bslexp}) 
and (\ref{bslth}).
However, as we lower the theoretical prediction for $B_{SL}$ by
increasing $\Gamma(b \rightarrow c\bar{c}s)$,
we simultaneously increase the prediction for $n_c$.
The theoretical prediction for $n_c$ obtained from eq.(\ref{bslth}) reads
\begin{equation}
\label{ncth}
n_c = 1.34 \mp 0.06
\end{equation}
The discrepancy between (\ref{ncexp}) and (\ref{ncth}) 
constitutes the "missing charm puzzle".

Furthermore CLEO collaboration reported rather large branching ratio of
$B \rightarrow \eta^\prime X_s$ decay mode \cite{Browder98}:
\begin{equation}
B(B \rightarrow \eta^\prime X_s) = (6.2 \pm 1.6 \pm 1.3) \times 10^{-4}
~~{\rm for}~ 2.0 < P_{\eta^\prime} <2.7 GeV
\end{equation}

These anomalies may be solved through an enhancement of the chromomagnetic
dipole coefficient $C_g$ by new physics.
For example, Lenz $et. al$ \cite{LNO97} showed that $|C_g (M_W)|$ must be
enhanced by a factor of 9 to 16 in order to explain the observed charm
deficit if the CKM structure of new physics contribution is the same as
in the Standard Model.

It is straightforward to include such effects of the enhanced chromomagnetic
moment $b\rightarrow s g$ on the nonleptonic two body decays of $B$ mesons.
This is accomplished by a simple replacement of
\begin{equation}
C_{g} \rightarrow \tilde{C_{g}} \equiv C_{g,SM} + C_{g,new}. 
\end{equation}
Here, to see the possible effects on the nonleptonic two body decays of
$B$ meson, we take two possibilities :
\begin{equation}
\tilde{C_{g}}(2.5 GeV) = + 5~ C_{g,SM}(2.5 GeV),
~~{\rm and}~~ - 5~ C_{g,SM}(2.5 GeV)
\end{equation}

In Fig. 7 and Fig. 8, we show the predictions for $B\rightarrow \pi K$ and
$B\rightarrow \eta^{'} K$ with $C_{g, new} = \pm 5 C_{g,SM}$ at
$\mu = 2.5$ GeV. The agreement between the data and the factorization 
predictions is generally degraded when one includes the enhanced 
$b \rightarrow s g$.  For the positive $C_{g,new} = +5 C_{g,SM}$, all the 
predictions become worse because of destructive interference between the 
SM amplitude and the enhanced $b\rightarrow s g$. For the negative $C_{g,new} 
= -5 C_{g,SM}$, one can improve the branching ratio for $B\rightarrow \eta^{'}
K$, only by paying a price to the worse predictions for $B \rightarrow \pi K$.
All these observations are based on the generalized factorization assumption.
In such an approximation, there is a tendency that the enhanced $b\rightarrow 
s g$ makes worse the agreements between the data and predictions. 
It has to be kept in mind that the enhanced $b\rightarrow sg$ scenario should 
be also tested in the exclusive $B$ decays in a better way than considered
in this work, if possible. 


\section{Conclusions and Discussions}

In this work, we considered the branching ratios and $CP$ asymmetries in $B$ 
meson decays into two light pseudoscalar mesons, and the possible effects
of the enhanced $b\rightarrow s g$ vertex suggested as a solution to the 
semileptonic branching ratio problem and the missing charm puzzle in $B$ 
meson decays. 
The typical branching ratios in various parameter set are given in 
Table 1. Their $\xi$ parameter dependences are given in Figs.~ 1, 2 and 3.
In $B \rightarrow \pi \pi, \pi K, K K$ modes, we included the EW penguin
effects as well as gluonic penguin effects. 
Especially in $B_d \rightarrow \pi^0 \pi^0, \pi^0 K^0, K^0 \eta$ modes and
$B^{\pm} \rightarrow \pi^0 K^{\pm}, K^{\pm} \eta$ modes,
EW penguin effects give 
important contributions to the branching ratio: about $+33 \%, +25 \%,
-36 \%, +29 \%, -38 \%$ variation respectively. 
In $B_d \rightarrow \pi^0 \pi^0, \eta^{(')} \eta^{(')}$ modes, 
the branching ratio plot has minimum value between 
$\xi=0.2$ and $\xi=0.3$ and its values become very small in this region
and very different from the values outside such the region.

The EW penguin effects and
the $b \rightarrow d ( s ) [ c \bar{c} \rightarrow g g
\rightarrow \eta^{(')}]$ type corrections on $CP$ violation
generally give small contributions to $CP$ asymmetry. The typical values of 
the $CP$ asymmetry of $B$ meson decays are given in Table 2. In Figs.~ 4, 5 
and 6, the $\xi$ dependences of $CP$ asymmetry in several modes are given in 
three different set of CKM angle $\rho, \eta$. For a fixed $\xi$, 
the important corrections of $CP$ asymmetry are given by 
varying the CKM angle itself.  The $\xi$ dependece of $CP$ asymmetry in 
charged $B$ decays are rather mild. In the neutral $B$ meson decay modes,
$CP$ asymmetry of $B_d \rightarrow \pi^0 \pi^0$ and 
$B_d \rightarrow \eta^{(')} \eta^{(')}$ modes change very sharply in the 
region  $0.2 < \xi < 0.4$. In other modes, the $CP$ asymmetry changes slowly 
when one varies $\xi$. 

Lastly, we made an observation that the enhanced $b\rightarrow s g$ tends 
to degrade the agreement between the data and the factorization predictions 
for $B$ meson decays into two light mesons, 
which seems to disfavor the enhanced $b\rightarrow s g$ scenario. 
The enhanced $b\rightarrow s g$ idea was put forward within the inclusive 
$B$ decays \cite{hou}, which can be studied with much less
theoretical uncertainties compared to the 
exclusive cases.
But it is certainly true that the enhanced $b\rightarrow s g$  
will affect the individual exclusive $B$ decay modes as well.
It would be very welcome to  study the 
effect of this enhanced $b\rightarrow s g$ on the individual exclusive $B$
decays in a way more reliable than the generalized factorization method 
employed in the present work. 


\section{Appendix}

In this appendix, we present the complete matrix elements including
the EW penguin effects and $f_{\eta^{(')}}^{(c)}$ effects.

\begin{itemize}
\item $B^{-} \rightarrow \pi^- \pi^0$
\begin{eqnarray}
M&=& \frac{G_F}{\sqrt{2}} \left\{ V_{ub} V_{ud}^* ( a_1 + a_2 )
  - V_{tb} V_{td}^* \left[ \frac{3}{2} ( a_9 - a_7 +a_{10} )
 +\frac{ 2 m_{\pi}^2 a_8 }{( m_u + m_d )( m_b - m_u )}\right. \right. 
 \nonumber \\
&& \left. \left. +\frac{  m_{\pi}^2 a_8 }{ 2 m_d ( m_b - m_u )} 
\right] \right\} 
 < \pi^- | \bar{d} u_{-} | 0 > < \pi^0 | \bar{u} b_{-} | B^{-} >
\end{eqnarray}
where
\begin{equation}
< \pi^- | \bar{d} u_{-} | 0 > < \pi^0 | \bar{u} b_{-} | B^{-} >
= i~ \frac{f_{\pi}}{\sqrt{2}}
( m_B^2 - m_{\pi}^2 ) F_0^{ B \rightarrow \pi}
( m_{\pi}^2 ).
\end{equation}

\item $\bar{B}^0 \rightarrow \pi^{+} \pi^{-}$

\begin{eqnarray}
M&=& \frac{G_F}{\sqrt{2}} \left\{ V_{ub} V_{ud}^*  a_1 
- V_{tb} V_{td}^* \left[ a_4 + a_{10}
+\frac{ 2 m_{\pi}^2 ( a_6 + a_8 )}{( m_u + m_d )( m_b - m_u )}
\right] \right\} \nonumber \\
&& \times
< \pi^{-} | \bar{d} u_{-} | 0 >< \pi^+ | \bar{u} b_{-} | \bar{B}^0 >
\end{eqnarray}
where
\begin{equation}
< \pi^{-} | \bar{d} u_{-} | 0 >< \pi^+ | \bar{u} b_{-} | \bar{B}^0 >
= i~ f_{\pi} ( m_B^2 - m_{\pi}^2 ) F_0^{ B \rightarrow \pi}
( m_{\pi}^2 )
\end{equation}

\item $\bar{B}^0 \rightarrow \pi^0 \pi^0$

\begin{eqnarray}
M&=& \frac{2 G_F}{\sqrt{2}}  \left\{  V_{ub} V_{ud}^*  a_2
+ V_{tb} V_{td}^* \left[  a_4 + \frac{3}{2} ( a_7 - a_9 ) 
- \frac{1}{2} a_{10} 
+ \frac{ m_{\pi}^2 ( 2 a_6 - a_8 )}{2 m_d ( m_b -m_d )}
\right] \right\}
 \nonumber \\ 
&& \times
< \pi^0 | \bar{u}u_{-} | 0 >< \pi^0 | \bar{d} b_{-} | \bar{B}^0 >
\end{eqnarray}
where
\begin{equation}
< \pi^0 | \bar{u}u_{-} | 0 >< \pi^0 | \bar{d} b_{-} | \bar{B}^0 >
= i~ \frac{f_{\pi}}{2} ( m_B^2 - m_{\pi}^2 )
F_0^{B \rightarrow \pi} ( m_{\pi}^2 ).
\end{equation}

\item $ B^- \rightarrow \bar{K}^0 \pi^-$

\begin{equation}
M= -\frac{G_F}{\sqrt{2}}  V_{tb} V_{ts}^*
\left[ a_4 - \frac{1}{2} a_{10}
+\frac{m_K^2 ( 2 a_6 - a_8 )}{( m_d + m_s )( m_b - m_d )}
\right]  
< \bar{K}^0 | \bar{s} d_{-} | 0 >< \pi^- | \bar{d}b_{-} | B^- >
\end{equation}
where
\begin{equation}
< \bar{K}^0 | \bar{s} d_{-} | 0 >< \pi^- | \bar{d}b_{-} | B^- >
= - i~ f_K ( m_B^2 - m_{\pi}^2 ) F_0^{B \rightarrow \pi}
( m_K^2 ).
\end{equation}

\item $B^- \rightarrow K^- \pi^0$

\begin{eqnarray}
M&=& \frac{G_F}{\sqrt{2}} \left\{ V_{ub} V_{us}^* 
( a_1 +  a_2 R_{K^- \pi^0} )
- V_{tb} V_{ts}^* \left[ a_4 + a_{10}
+\frac{2 m_K^2 ( a_6 + a_8)}{( m_u + m_s )( m_b - m_u )}
\right. \right. \nonumber \\
&& \left. \left. +\frac{3}{2} ( a_9 - a_7 ) R_{K^- \pi^0}  
\right] \right\}
< K^- | \bar{s} u_{-} | 0 >< \pi^0 | \bar{u} b_{-} | B^- >
\end{eqnarray}
where
\begin{eqnarray}
&&R_{K^- \pi^0} \equiv
\frac{< \pi^0 | \bar{u} u_{-} | 0 > < K^- | \bar{s} b_{-} | B^- >}
     {< K^- | \bar{s} u_{-} | 0 >< \pi^0 | \bar{u} b_{-} | B^- >}=
\frac{ f_{\pi}}{f_K} \frac{m_B^2-m_K^2}{m_B^2-m_{\pi}^2}
\frac{F_0^{B \rightarrow K^-} ( m_{\pi}^2 )}
     {F_0^{B \rightarrow \pi^0} ( m_K^2 )}, \nonumber \\
&&< K^- | \bar{s} u_{-} | 0 >< \pi^0 | \bar{u} b_{-} | B^- >
= i~ \frac{f_K}{\sqrt{2}} ( m_B^2 - m_{\pi}^2 ) F_0^{B \rightarrow \pi}
( m_K^2 ).
\end{eqnarray}

\item $\bar{B}^0 \rightarrow K^- \pi^+$

\begin{eqnarray}
M&=& \frac{G_F}{\sqrt{2}} \left\{ V_{ub} V_{us}^*~ a_1
- V_{tb} V_{ts}^* \left[ a_4 + a_{10} +
\frac{2 m_K^2 ( a_6 + a_8 )}{( m_u + m_s )( m_b - m_u )}
\right] \right\} \nonumber \\
&& \times
< K^- | \bar{s} u_{-} | 0 >< \pi^+ | \bar{u} b_{-} | \bar{B}^0 >
\end{eqnarray}
where
\begin{equation}
< K^- | \bar{s} u_{-} | 0 >< \pi^+ | \bar{u} b_{-} | \bar{B}^0 >
= i~ f_K ( m_B^2 - m_{\pi}^2 ) F_0^{B \rightarrow \pi} ( m_K^2 ).
\end{equation}

\item $\bar{B}^0 \rightarrow \bar{K}^0 \pi^0$

\begin{eqnarray}
M&=& \frac{G_F}{\sqrt{2}} \left\{ V_{ub} V_{us}^* a_2 R_{\bar{K}^0 \pi^0}
- V_{tb} V_{ts}^* \left[ a_4 - \frac{1}{2} a_{10}
+ \frac{m_K^2 ( 2 a_6 - a_8 )}{( m_d + m_s )( m_b - m_d )}
\right. \right. \nonumber \\
&& \left. \left. + \frac{3}{2} ( a_9 - a_7 ) R_{\bar{K}^0 \pi^0}
\right] \right\}
< \bar{K}^0 | \bar{s} d_{-} | 0 >< \pi^0 | \bar{d} b_{-} | \bar{B}^0 >
\end{eqnarray}
where
\begin{eqnarray}
&& R_{\bar{K}^0 \pi^0} \equiv
\frac{< \pi^0 | \bar{u} u_{-} | 0 >< \bar{K}^0 | \bar{s} b_{-} | \bar{B}^0 >}
     {< \bar{K}^0 | \bar{s} d_{-} | 0 >< \pi^0 | \bar{d} b_{-} | \bar{B}^0 >}
=- \frac{ f_{\pi}}{f_K} \frac{m_B^2-m_K^2}{m_B^2-m_{\pi}^2}
\frac{F_0^{B \rightarrow K} ( m_{\pi}^2 )}
     {F_0^{B \rightarrow \pi} ( m_K^2 )}, \nonumber \\
&&< \bar{K}^0 | \bar{s} d_{-} | 0 >< \pi^0 | \bar{d} b_{-} | \bar{B}^0 >
= - i~ \frac{f_K}{\sqrt{2}}
 ( m_B^2 - m_K^2 ) F_0^{B \rightarrow \pi} ( m_K^2 ).
\end{eqnarray}

\item $B^{-} \rightarrow K^0 K^{-}$

\begin{eqnarray}
M&=& -\frac{G_F}{\sqrt{2}} V_{tb} V_{td}^*
\left\{ a_4 - \frac{1}{2} a_{10} 
+ \frac{m_K^2 ( 2 a_6 - a_8 )}{( m_d + m_s )( m_b - m_s )}
\right\} \nonumber \\
&& \times
< K^0 | \bar{d} s_{-} | 0 >< K^{-} | \bar{s} b_{-} | B^{-} >
\end{eqnarray}
where
\begin{equation}
< K^0 | \bar{d} s_{-} | 0 >< K^{-} | \bar{s} b_{-} | B^{-} >
= - i~ f_K ( m_B^2 - m_K^2 ) F_0^{B \rightarrow K} ( m_K^2 ).
\end{equation}

\item $\bar{B}^0 \rightarrow K^0 \bar{K}^0$

\begin{eqnarray}
M&=& -\frac{G_F}{\sqrt{2}} V_{tb} V_{td}^*
\left\{ a_4 - \frac{1}{2} a_{10} 
+ \frac{m_K^2 ( 2 a_6 - a_8 )}{( m_d + m_s )( m_b - m_s )}
\right\} \nonumber \\
&& \times
< K^0 | \bar{d} s_{-} | 0 >< \bar{K}^0 | \bar{s} b_{-} | \bar{B}^0 >
\end{eqnarray}
where
\begin{equation}
< K^0 | \bar{d} s_{-} | 0 >< \bar{K}^0 | \bar{s} b_{-} | \bar{B}^0 >
= - i~ f_K ( m_B^2 - m_K^2 ) F_0^{B \rightarrow K} ( m_K^2 ).
\end{equation}

\item $B^- \rightarrow \pi^- \eta$

\begin{eqnarray}
M&=& \frac{G_F}{\sqrt{2}} \left\{ V_{ub} V_{ud}^*
 ~(~a_2 + a_1 ~R_{\pi^{-} \eta}~ ) + V_{cb} V_{cd}^*~a_2~\frac{f_{\eta}^{(c)}}
{f_{\eta}^u} \right. \nonumber \\
&&- V_{tb} V_{td}^* \left[~
2 ( a_3 - a_5 ) + a_4 
+\frac{m_{\eta}^2 ( 2 a_6 - a_8 )}{2 m_s ( m_b - m_s )}
\left( 1 - \frac{f_{\eta}^u}{f_{\eta}^s} \right)
- \frac{1}{2} ( a_7 - a_9 + a_{10} )
\right. \nonumber \\
&&+\left\{ a_4 + a_{10} + \frac{ 2 m_{\pi}^2 ( a_6 + a_8 )}
{( m_u + m_d )( m_b - m_u )} \right\} R_{\pi^{-} \eta}  \nonumber \\
&&+ \left. \left. \left\{ a_3 - a_5 + \frac{1}{2} ( a_7 - a_9 )
                 \right\} \frac{f_{\eta}^s}{f_{\eta}^u}
+\left\{ a_3 - a_5 - a_7 + a_9  \right\}
\frac{f_{\eta}^{(c)}}{f_{\eta}^u} \right] \right\} \nonumber \\
&&\times < \pi^{-} | \bar{d} b_{-} | \ B^{-} >
< \eta | \bar{u} u_{-} | 0 >
\end{eqnarray}
where
\begin{eqnarray}
\label{matrixetapi2}
&&R_{\pi^{-} \eta} \equiv
\frac{< \pi^{-} | \bar{u} u_{-} | 0 > < \eta | \bar{d} b_{-} | B^- >}
     {< \eta | \bar{u} u_{-} | 0 > < \pi^{-} | \bar{d} b_{-} | B^- >} =
\frac{f_{\pi}}{f_{\eta}^u }~
\frac{ m_B^2 - m_{\eta}^2 }{ m_B^2 - m_{\pi}^2 }~
\frac{ F_0^{B \rightarrow  \eta} ( m_{\pi}^2 )}
     { F_0^{ B \rightarrow \pi} ( m_{\eta}^2 ) },
\nonumber \\
&&< \pi^{-} | \bar{d} b_{-} | \ B^{-} > < \eta | \bar{u} u_{-} | 0 > =
i~ f_{\eta}^u~ ( m_B^2 - m_{\pi}^2 )~ F_0^{B \rightarrow \pi }
( m_K^2 ).
\end{eqnarray}

\item $\bar{B}^0 \rightarrow \pi^0 \eta$

\begin{eqnarray}
M&=& \frac{G_F}{\sqrt{2}} \left\{ V_{ub} V_{ud}^*~ a_2 ~( 1 + R_{\pi^0 \eta} )
 + V_{cb} V_{cd}^*~ a_2~ \frac{f_{\eta}^{(c)}}{f_{\eta}^{u}} \right.
\nonumber \\
&& - V_{tb} V_{td}^* \left[ 2 ( a_3 - a_5 ) +  a_4 
+\frac{m_{\eta}^2 ( 2 a_6 - a_8 )}{2 m_s ( m_b - m_s )}
\left( 1 - \frac{f_{\eta}^u}{f_{\eta}^s} \right)
-\frac{1}{2} ( a_7  - a_9 + a_{10} )
\right. \nonumber \\
&&- \left\{ a_4 + ( a_6 - \frac{1}{2} a_8 ) 
\frac{ m_{\pi}^2}{m_d ( m_b - m_d )} 
+ \frac{3}{2} ( a_7 - a_9 ) 
- \frac{1}{2}  a_{10} \right\}~ R_{\pi^0 \eta}
\nonumber \\
&& \left. \left.+\left\{ a_3 - a_5 + \frac{1}{2} ( a_7 - a_9 ) \right\} 
 \frac{f_{\eta}^s}{f_{\eta}^u}
+ \left\{ a_3 - a_5 - a_7 + a_9 ) \right\} 
  \frac{f_{\eta}^{(c)}}{f_{\eta}^u} \right] \right\}
\nonumber \\
&&\times < \eta | \bar{u} u_{-} | 0 >
< \pi^0 | \bar{d} b_{-} | \bar{B}^0 >
\end{eqnarray}
where
\begin{eqnarray}
&&R_{\pi^0 \eta} \equiv 
\frac{< \pi^0 | \bar{u} u_{-} | 0 > < \eta | \bar{d} b_{-} | \bar{B}^0 >}
     {< \eta | \bar{u} u_{-} | 0 > < \pi^0 | \bar{d} b_{-} | \bar{B}^0 >} =
- \frac{f_{\pi}}{f_{\eta}^u }~ 
\frac{ m_B^2 - m_{\eta}^2 }{ m_B^2 - m_{\pi}^2 }~
\frac{ F_0^{B \rightarrow  \eta} ( m_{\pi}^2 )}
     { F_0^{ B \rightarrow \pi} ( m_{\eta}^2 ) }, 
\nonumber \\
&&< \eta | \bar{u} u_{-} | 0 > < \pi^0 | \bar{d}
b_{-} | \bar{B}^0 > = i ~\frac{f_{\eta}^u}{\sqrt{2}} ~( m_B^2 - m_{\pi}^2 )~ 
F_0^{B \rightarrow \pi} ( m_{\eta}^2 ).
\end{eqnarray}

\item The matrix elements of the 
$B^- \rightarrow \pi^- \eta', \bar{B}^0 \rightarrow \pi^0 \eta'$ modes
might be obtained by replacing  $\eta$ with $\eta'$ in
$B^- \rightarrow \pi^- \eta, \bar{B}^0 \rightarrow \pi^0 \eta$ modes.

\item $B^- \rightarrow K^- \eta$

\begin{eqnarray}
M &=& \frac{G_F}{\sqrt{2}} \left\{ V_{ub} V_{us}^*~ ( a_2 + 
a_1 R_{K^- \eta} ) + V_{cb} V_{cs}^*~ a_2 \frac{f_{\eta}^{(c)}}{f_{\eta}^u}
- V_{tb} V_{ts}^* \left[ 2 ( a_3 - a_5 ) -
\frac{ m_{\eta}^2 ( 2 a_6 - a_8 )}{2 m_s ( m_b - m_s )}
\right. \right.  \nonumber \\
&& - \frac{1}{2} ( a_7 - a_9 ) +
\left\{ a_4 + \frac{2 m_K^2 ( a_6 + a_8 )}{( m_s + m_u )( m_b - m_u )}
+ a_{10} \right\} R_{K^- \eta}  \nonumber \\
&& + \left\{ a_3 + a_4 - a_5 + 
\frac{ m_{\eta}^2 ( 2 a_6 - a_8 )}{ 2 m_s ( m_b - m_s )} 
+ \frac{1}{2} ( a_7 - a_9 - a_{10} ) \right\} \frac{f_{\eta}^s}{f_{\eta}^u}
\nonumber \\
&& \left. \left.
+ ( a_3 - a_5 - a_7 + a_9 ) \frac{f_{\eta}^{(c)}}{f_{\eta}^u}
\right] \right\}
< \eta | \bar{u} u_{-} | 0 >< K^- | \bar{s} b_{-} | B^- >
\end{eqnarray}
where
\begin{eqnarray}
&&R_{K^- \eta} \equiv 
\frac{< K^- | \bar{s} u_{-} | 0 > < \eta | \bar{u} b_{-} | B^- >}
     {< \eta | \bar{u} u_{-} | 0 > < K^- | \bar{s} b_{-} | B^- >} =
\frac{f_K}{f_{\eta}^u }~ 
\frac{ m_B^2 - m_{\eta}^2 }{ m_B^2 - m_K^2 }~
\frac{ F_0^{B \rightarrow  \eta} ( m_K^2 )}
     { F_0^{ B \rightarrow K} ( m_{\eta}^2 ) }, 
\nonumber \\
&&< \eta | \bar{u} u_{-} | 0 > < K^- | \bar{d} b_{-} | B^- > 
= i ~f_{\eta}^u ~( m_B^2 - m_K^2 )~ 
F_0^{B \rightarrow K} ( m_{\eta}^2 ).
\end{eqnarray}
\item $\bar{B}^0 \rightarrow \bar{K}^0 \eta$
\begin{eqnarray}
M &=& \frac{G_F}{\sqrt{2}} \left\{ V_{ub} V_{us}^*~  a_2 + 
 V_{cb} V_{cs}^*~ a_2 \frac{f_{\eta}^{(c)}}{f_{\eta}^u}
- V_{tb} V_{ts}^* \left[ 2 ( a_3 - a_5 ) -
\frac{ m_{\eta}^2 ( 2 a_6 - a_8 )}{2 m_s ( m_b - m_s )}
 - \frac{1}{2} ( a_7 - a_9 ) 
\right. \right.  \nonumber \\
&& + \left\{ a_4 + \frac{ m_K^2 ( 2 a_6 - a_8 )}{( m_s + m_d )( m_b - m_d )}
- \frac{1}{2} a_{10} \right\} R_{\bar{K}^0 \eta}  \nonumber \\
&& + \left\{ a_3 + a_4 - a_5 + 
\frac{ m_{\eta}^2 ( 2 a_6 - a_8 )}{ 2 m_s ( m_b - m_s )} 
+ \frac{1}{2} ( a_7 - a_9 - a_{10} ) 
\right\} \frac{f_{\eta}^s}{f_{\eta}^u}
\nonumber \\
&& \left. \left.
+ ( a_3 - a_5 - a_7 + a_9 ) \frac{f_{\eta}^{(c)}}{f_{\eta}^u}
\right] \right\}
< \eta | \bar{u} u_{-} | 0 >< \bar{K}^0 | \bar{s} b_{-} | \bar{B}^0 >
\end{eqnarray}
where
\begin{eqnarray}
&&R_{\bar{K}^0 \eta} \equiv 
\frac{< \bar{K}^0 | \bar{s} d_{-} | 0 >
          < \eta | \bar{d} b_{-} | \bar{B}^0 >}
     {< \eta | \bar{u} u_{-} | 0 > 
          < \bar{K}^0 | \bar{s} b_{-} | \bar{B}^0 >} =
\frac{f_K}{f_{\eta}^u }~ 
\frac{ m_B^2 - m_{\eta}^2 }{ m_B^2 - m_K^2 }~
\frac{ F_0^{B \rightarrow  \eta} ( m_K^2 )}
     { F_0^{ B \rightarrow K} ( m_{\eta}^2 ) }, 
\nonumber \\
&&< \eta | \bar{u} u_{-} | 0 > < \bar{K}^0 | \bar{s} b_{-} | \bar{B}^0 >
= i ~f_{\eta}^u ~( m_B^2 - m_K^2 )~ 
F_0^{B \rightarrow K} ( m_{\eta}^2 ).
\end{eqnarray}

\item The matrix elements of the 
$B^- \rightarrow K^- \eta', \bar{B}^0 \rightarrow \bar{K}^0 \eta'$ modes
might be obtained by replacing  $\eta$ with $\eta'$ in
$B^- \rightarrow K^- \eta, \bar{B}^0 \rightarrow \bar{K}^0 \eta$ modes.
\item $\bar{B}^0 \rightarrow \eta \eta$

\begin{eqnarray}
M&=& \frac{2 G_F}{\sqrt{2}} \left\{ V_{ub} V_{ud}^*~ a_2 +
   V_{cb} V_{cd}^*~ a_2~ \frac{f_{\eta}^{(c)}}{f_{\eta}^u}  
\right. \\ \nonumber
&&- V_{tb} V_{td}^* \left[ 2 ( a_3 - a_5 ) + a_4 
+\frac{m_{\eta}^2 ( 2 a_6 - a_8 )}{2 m_s ( m_b - m_s )}
\left( 1 - \frac{f_{\eta}^u}{f_{\eta}^s} \right)
  - \frac{1}{2} ( a_7 - a_9 + a_{10} ) 
\right.  \nonumber \\
&& \left. \left.  + \left\{ a_3 - a_5 + \frac{1}{2} ( a_7 - a_9 ) \right\}
   \frac{f_{\eta}^s}{f_{\eta}^u}  
 + ( a_3 - a_5 - a_7 + a_9 ) \frac{f_{\eta}^{(c)}}{f_{\eta}^u} 
  \right] \right\} \nonumber \\ 
&& \times < \eta | \bar{u} u_{-} | 0 > < \eta | \bar{d} b_{-} | \bar{B}^0 >
\end{eqnarray}
where
\begin{equation}
< \eta | \bar{u} u_{-} | 0 > < \eta | \bar{d} b_{-} | \bar{B}^0 >
= - i ~ f_{\eta}^u~( m_B^2 - m_{\eta}^2 )~ F_0^{ B \rightarrow
\eta } ( m_{\eta}^2 ).
\end{equation}

\item The matrix elements of the 
$B^0 \rightarrow \eta' \eta'$
might be obtained by replacing  $\eta$ with $\eta'$ in
$B^0 \rightarrow \eta \eta$ modes.

\item $\bar{B}^0 \rightarrow \eta \eta'$

\begin{eqnarray}
M&=& \frac{G_F}{\sqrt{2}} \left\{ V_{ub} V_{ud}^*~ a_2 +
   V_{cb} V_{cd}^*~ a_2~ \frac{f_{\eta}^{(c)}}{f_{\eta}^u}  
\right. \\ \nonumber
&&- V_{tb} V_{td}^* \left[ 2 ( a_3 - a_5 ) + a_4 
+\frac{m_{\eta}^2 ( 2 a_6 - a_8 )}{2 m_s ( m_b - m_s )}
\left( 1 - \frac{f_{\eta}^u}{f_{\eta}^s} \right)
  - \frac{1}{2} ( a_7 - a_9 + a_{10} ) 
 \right.  \nonumber \\
&&
\left. \left. + \left\{ a_3 - a_5 + \frac{1}{2} ( a_7 - a_9 ) \right\}
   \frac{f_{\eta}^s}{f_{\eta}^u}  
+ ( a_3 - a_5 - a_7 + a_9 ) \frac{f_{\eta}^{(c)}}{f_{\eta}^u} 
  \right] \right\} 
< \eta | \bar{u} u_{-} | 0 > < \eta' | \bar{d} b_{-} | \bar{B}^0 > 
\nonumber \\
&&
+ ( \eta \rightarrow \eta' )
\end{eqnarray}
where
\begin{eqnarray}
< \eta | \bar{u} u_{-} | 0 > < \eta' | \bar{d} b_{-} | \bar{B}^0 >
&=& - i ~ f_{\eta}^u~( m_B^2 - m_{\eta'}^2 )~ F_0^{ B \rightarrow
\eta' } ( m_{\eta}^2 ), \nonumber \\
< \eta' | \bar{u} u_{-} | 0 > < \eta | \bar{d} b_{-} | \bar{B}^0 >
&=& - i ~ f_{\eta'}^u~( m_B^2 - m_{\eta}^2 )~ F_0^{ B \rightarrow
\eta } ( m_{\eta}^2 ).
\end{eqnarray}

\end{itemize}

\section{Acknowledgements}
We thank Ahmed Ali and Cai-Dian L\"{u} for useful correspondence for their 
works, hep-ph/9804363 and hep-ph/9805403 that helped us to fix some mistakes
in our original calculations.    
This work is supported in part by  KOSEF, Contract 971-0201-002-2, KOSEF  
through Center for Theoretical Physics at Seoul National University, 
the German-Korean scientific exchange programme DFG-446-KOR-113/72/0,
the Ministry of Education through the Basic Science Research Institute, 
Contract No. BSRI-96-2418 (PK), and by KRF postdoctoral program (YGK).

\newpage

\begin{table}
\caption{Combined branching ratios in unit of $10^{-5}$. 'QCD'
and 'EW' present the QCD penguin and EW penguin effects respectively.
'Full' includes the $b \rightarrow s [ c \bar{c} 
\rightarrow gg \rightarrow \eta \eta' ]$ effects :
$\xi=1/3$, $f_{\eta}^{(c)}= -0.9 MeV$, $f_{\eta'}^{(c)}= -2.3 MeV$
( $m_c = 1.5 GeV$ ), $\rho=0.05$, $\eta=0.36$ }
\begin{tabular}{l|c|c|c|c|c} 
Decay Mode& Tree& Tree+QCD& \multicolumn{2}{c|}{Tree+QCD+EW}& Exp. \\ \hline
$B^{\pm} \rightarrow \pi^{\pm} \pi^0$& 0.51&                0.51&
  \multicolumn{2}{c|}{0.51}&  $<$2.0  \\
$B^{\pm} \rightarrow \pi^{\pm} K^0$&  0&                   1.63&
   \multicolumn{2}{c|}{1.61}&   $2.3^{+1.1}_{-1.0}\pm 0.3 \pm 0.2$ \\
$B^{\pm} \rightarrow \pi^0 K^{\pm}$&   0.038&               0.84&
   \multicolumn{2}{c|}{1.08}&   $<$1.6 \\
$B^{\pm} \rightarrow  K^{\pm} K^0$&     0&                   0.082&
   \multicolumn{2}{c|}{0.081}&   $<$2.1 \\ \hline
$B_d \rightarrow \pi^{\pm} \pi^{\mp}$& 0.94&                0.93&
   \multicolumn{2}{c|}{0.93}&  $<$1.5  \\
$B_d \rightarrow \pi^0 \pi^0$&         $0.11\times 10^{-2}$&  0.021&
   \multicolumn{2}{c|}{0.014}& $<$0.93   \\
$B_d \rightarrow \pi^{\pm} K^{\mp}$&   0.071&               1.68&
   \multicolumn{2}{c|}{1.71}&   $1.5^{+0.5}_{-0.4} \pm 0.1 \pm 0.1$ \\
$B_d \rightarrow \pi^0 K^0$&           $0.53\times 10^{-4}$&  0.83&
   \multicolumn{2}{c|}{0.62}&   $<$4.1 \\
$B_d \rightarrow K^0 \bar{K}^0$&   0&                   0.082&
   \multicolumn{2}{c|}{0.081}&   $<$1.7 \\  \hline \hline
Decay Mode& Tree& Tree+QCD& Tree+QCD+EW& Full& Exp. \\ \hline
$B^{\pm} \rightarrow \pi^{\pm} \eta$&  0.23&                0.25&
  0.26& 0.26&  $<$1.5  \\
$B^{\pm} \rightarrow \pi^{\pm} \eta'$& 0.16&                0.18&
  0.17& 0.18&   $<$3.1 \\
$B_d \rightarrow \pi^0 \eta$&      $0.18\times 10^{-4}$& 0.026&
  0.026& 0.026& $<$0.8   \\
$B_d \rightarrow \pi^0 \eta'$&     $0.71\times 10^{-5}$& 0.011&
  $0.79 \times 10^{-2}$& $0.81 \times 10^{-2}$&    $<$1.1\\ \hline
$B^{\pm} \rightarrow K^{\pm} \eta$&   0.017&               0.18&
  0.11& 0.11&   $<$1.4 \\
$B^{\pm} \rightarrow K^{\pm} \eta'$&  0.012&               2.15&
  2.05& 2.08&  $6.5^{+1.5}_{-1.4} \pm 0.9$  \\  
$B_d \rightarrow K^0 \eta$&           $0.37\times 10^{-4}$&  0.15&
  0.096& 0.094&  $<$3.3  \\
$B_d \rightarrow K^0 \eta'$&          $0.24\times 10^{-4}$&  2.16&
  2.04& 2.07&  $4.7^{+2.7}_{-2.0} \pm 0.9$  \\ \hline
$B_d \rightarrow \eta \eta$&        $0.32\times 10^{-3}$& $0.83\times 10^{-2}$&
  0.010&  0.010&  $<$1.8  \\
$B_d \rightarrow \eta \eta'$&       $0.44\times 10^{-3}$& $0.78\times 10^{-2}$&
  $0.83 \times 10^{-2}$&  $0.86 \times 10^{-2}$&   $<$2.7 \\
$B_d \rightarrow \eta' \eta'$&      $0.15\times 10^{-3}$& $0.17\times 10^{-3}$&
  $0.14 \times 10^{-3}$&  $0.15 \times 10^{-3}$&  $<$4.7\\ 
\end{tabular}
\end{table}


\begin{table}
\caption{$CP$ asymmetries in $\%$.
'QCD' and 'EW' present the QCD penguin and EW penguin effects respectively.
'Full' includes the $b \rightarrow s [ c \bar{c} 
\rightarrow gg \rightarrow \eta \eta' ]$ effects.
'Full(one-mixing)' presents the one-mixing scheme for
$\eta-\eta'$ mixing :
CKM phases are $\rho = 0.05, \eta = 0.36 ( \rho = 0.30, \eta = 0.42;
\rho = 0, \eta = 0.22),  \xi = 1/3$}
\begin{tabular}{l|c|c|c|c} 
Decay Mode& QCD& \multicolumn{3}{c}{QCD+EW} \\ \hline
$B^{\pm} \rightarrow \pi^{\pm} \pi^0$&  0 ( 0, 0)&    
\multicolumn{3}{c}{$-0.05 ( -0.03, -0.09)$}\\
$B^{\pm} \rightarrow \pi^{\pm} K^0$&    1.6 ( 1.8, 1.0)&  
\multicolumn{3}{c}{1.6 ( 1.8, 1.0)}\\
$B^{\pm} \rightarrow \pi^0 K^{\pm}$&   8.4 ( 12.9, 5.0)&   
\multicolumn{3}{c}{6.7 ( 10.0, 4.0)}\\
$B^{\pm} \rightarrow K^0 K^{\pm}$&   $ -12.2 ( -20.5, -7.3)$ &  
\multicolumn{3}{c}{$-12.3 ( -20.6, -7.3)$}\\
\hline
$B_d \rightarrow \pi^{\pm} \pi^{\mp}$& 5.1 ( 21.4, 22.8)&   
\multicolumn{3}{c}{5.4 ( 21.5, 23.3)}\\
$B_d \rightarrow \pi^0 \pi^0$&      $  -10.5 ( -18.8, -6.5)$ &  
\multicolumn{3}{c}{$-13.9 ( -24.4, -8.6)$}\\
$B_d \rightarrow \pi^0 K_S$&           31.1 ( 41.1 19.8)&  
\multicolumn{3}{c}{31.1 ( 41.1, 19.8)}\\
$B_d \rightarrow K_S K_S$&      15.2 ( 26.0, 9.1)&  
\multicolumn{3}{c}{15.2 ( 26.1, 9.1)}\\ \hline \hline
Decay Mode & QCD & QCD+EW & Full & Full(one-mixing)  \\ \hline
$B^{\pm} \rightarrow \pi^{\pm} \eta$& $-18.2 ( -9.6, -23.5)$ & 
 $-18.0 ( -9.6, -22.4)$ & $-18.0 ( -9.6, -22.4)$ & 
$-18.6 ( -9.9, -23.1)$ \\
$B^{\pm} \rightarrow \pi^{\pm} \eta'$& $-17.9 ( -9.4, -25.9)$&
 $-18.0 ( -9.4, -26.3)$& $-17.9 ( -9.4, -26.1)$&
 $-17.7 ( -9.3, -25.9)$ \\
$B^{\pm} \rightarrow K^{\pm} \eta$&   $ -6.7 ( -5.3, -4.8)$ &
 $-10.9 ( -7.9, -8.2)$ & $-11.1 ( -8.0, -8.4)$ &
 $-13.4 ( -7.5, -14.5)$ \\
$B^{\pm} \rightarrow K^{\pm} \eta'$&   4.5 ( 5.8, 2.7)&
   4.7 ( 6.1, 2.8)& 4.6 ( 6.0, 2.8)& 
4.8 ( 6.3, 2.9)\\
\hline
$B_d \rightarrow \pi^0 \eta$&        $  18.8 ( 31.4, 11.3)$ & 
$18.8 ( 31.4, 11.3)$ & $18.9 ( 31.6, 11.3)$ & 
$19.2 ( 32.0, 11.5)$ \\
$B_d \rightarrow \pi^0 \eta'$&        $ 23.5 ( 38.4, 14.2)$ &  
$27.9 ( 44.5, 16.8)$ & $28.1 ( 44.8, 17.0)$ &
 $28.1 ( 44.7, 17.0)$  \\
$B_d \rightarrow K_S \eta$&            31.8 ( 41.7, 20.3)&  
32.0 ( 41.9, 20.5)& 32.1 ( 41.9, 20.5)& 
34.3 ( 43.6, 22.1)\\
$B_d \rightarrow K_S \eta'$&           29.9 ( 40.1, 19.0)&  
29.9 ( 40.0, 18.9)& 29.9 ( 40.1, 18.9)&
 29.9 ( 40.0, 18.9)\\
$B_d \rightarrow \eta \eta$&          37.3 ( 52.4, 22.9)& 
33.1 ( 48.1, 20.2)& 33.1 ( 48.1, 20.2)& 
33.5 ( 48.6, 20.4)\\
$B_d \rightarrow \eta \eta'$&         45.5 ( 59.6, 28.5)&
43.9 ( 58.3, 27.4)& 43.8 ( 58.1, 27.4)&
43.9 ( 58.3, 27.4)\\
$B_d \rightarrow \eta' \eta'$&        57.7 ( 67.8, 37.8)&
63.1 ( 70.6, 42.5)& 62.4 ( 69.9, 41.9)& 
62.5 ( 69.8, 42.0)\\
\end{tabular}
\end{table}


\begin{figure}
\centerline{\epsffile{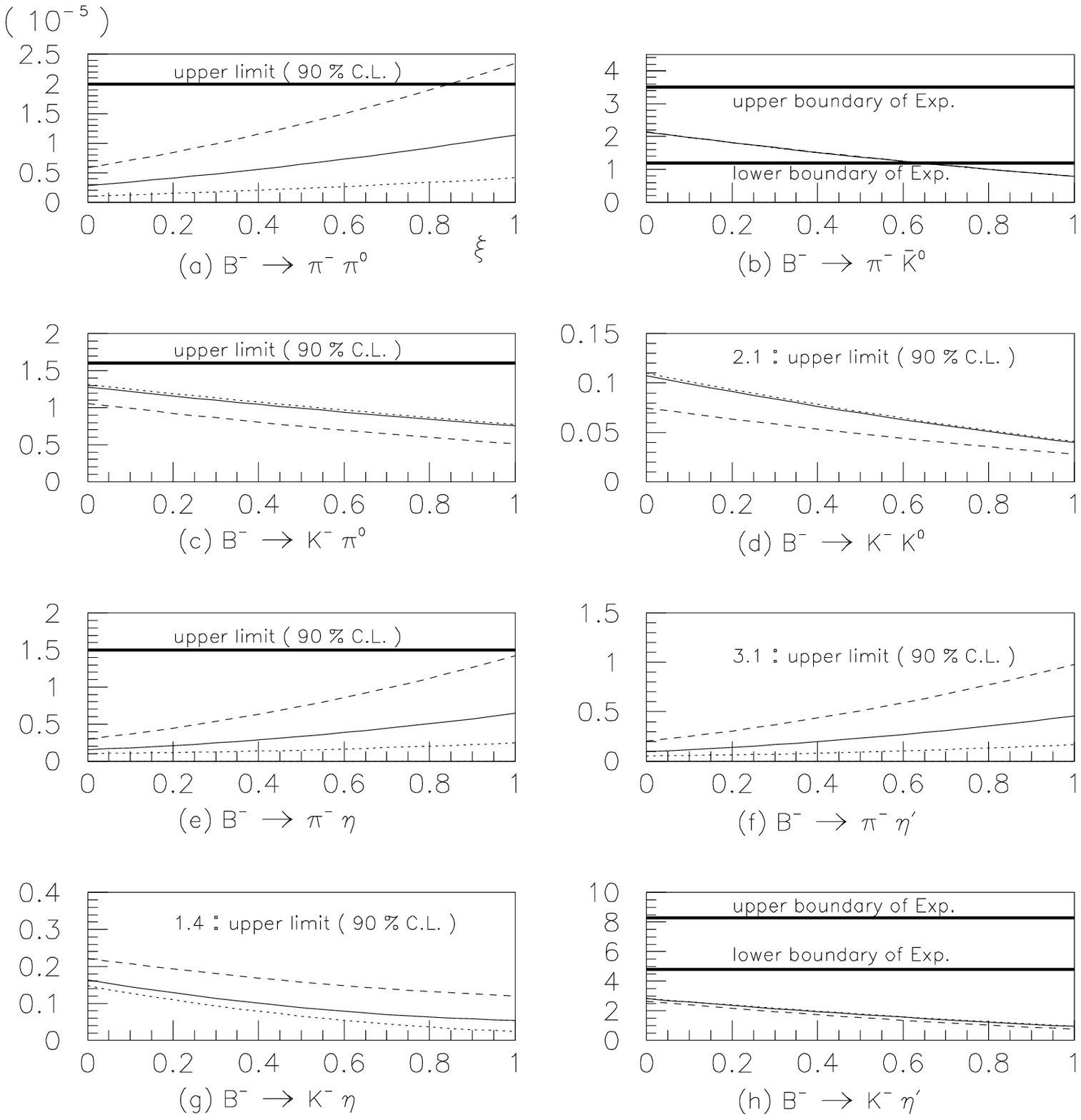}}
\caption{ Combined branching ratio of the charged $B$ meson decays as
a function of nonfactorization parameter $\xi$: 
$\rho = 0.05, \eta = 0.36$ for real line, 
$\rho = 0.30, \eta = 0.42$ for dashed line and
$\rho = 0, \eta = 0.22$ for dotted line. }
\label{fig1}
\end{figure}

\begin{figure}
\centerline{\epsffile{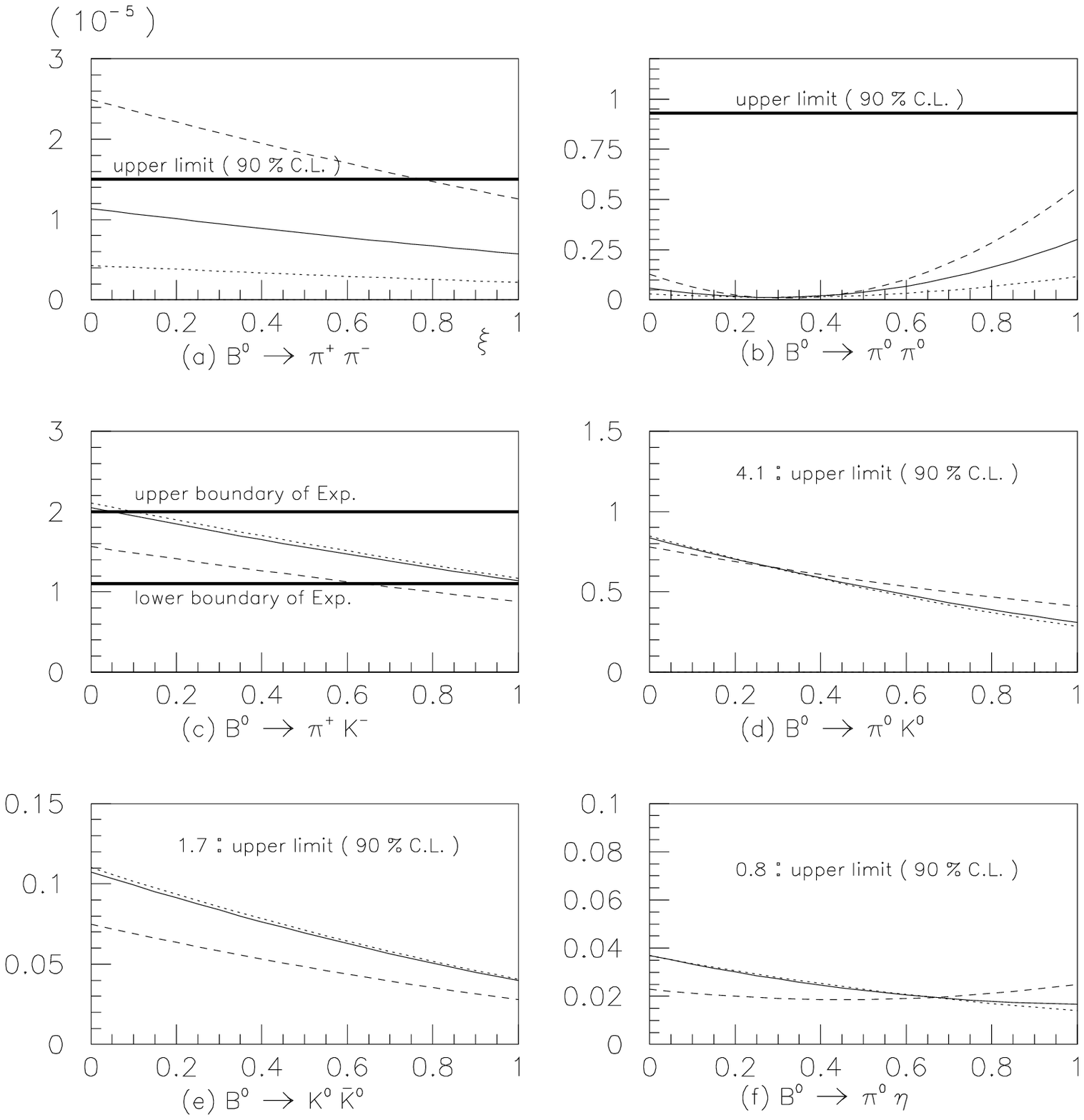}}
\caption{ Combined branching ratio of the neutral $B$ meson decays as
a function of nonfactorization parameter $\xi$: 
$\rho = 0.05, \eta = 0.36$ for real line, 
$\rho = 0.30, \eta = 0.42$ for dashed line and
$\rho = 0, \eta = 0.22$ for dotted line. }
\label{fig2}
\end{figure}

\begin{figure}
\centerline{\epsffile{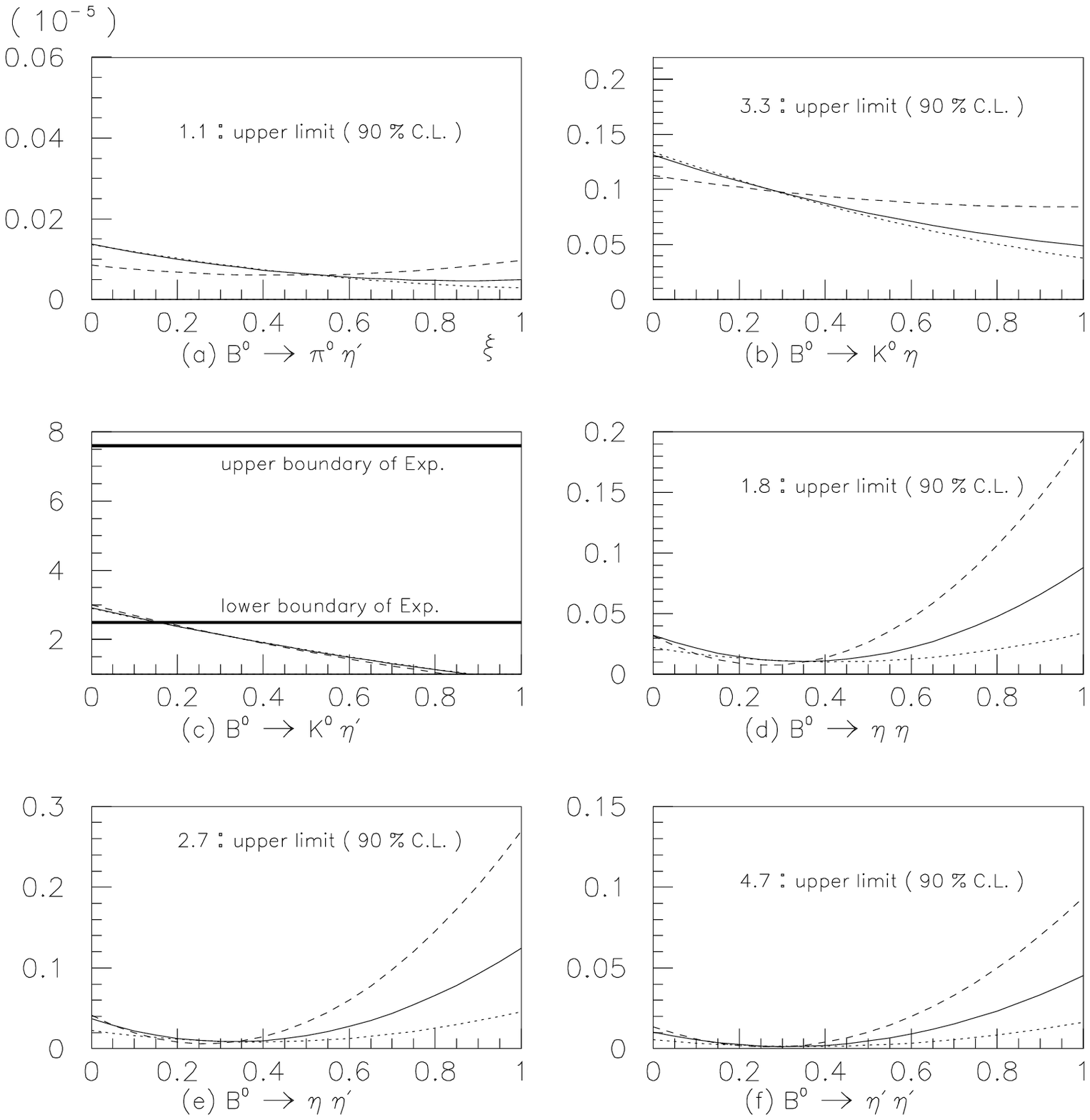}}
\caption{ Combined branching ratio of the neutral $B$ meson decays as
a function of nonfactorization parameter $\xi$: 
$\rho = 0.05, \eta = 0.36$ for real line, 
$\rho = 0.30, \eta = 0.42$ for dashed line and
$\rho = 0, \eta = 0.22$ for dotted line. }
\label{fig2-1}
\end{figure}

\begin{figure}
\centerline{\epsffile{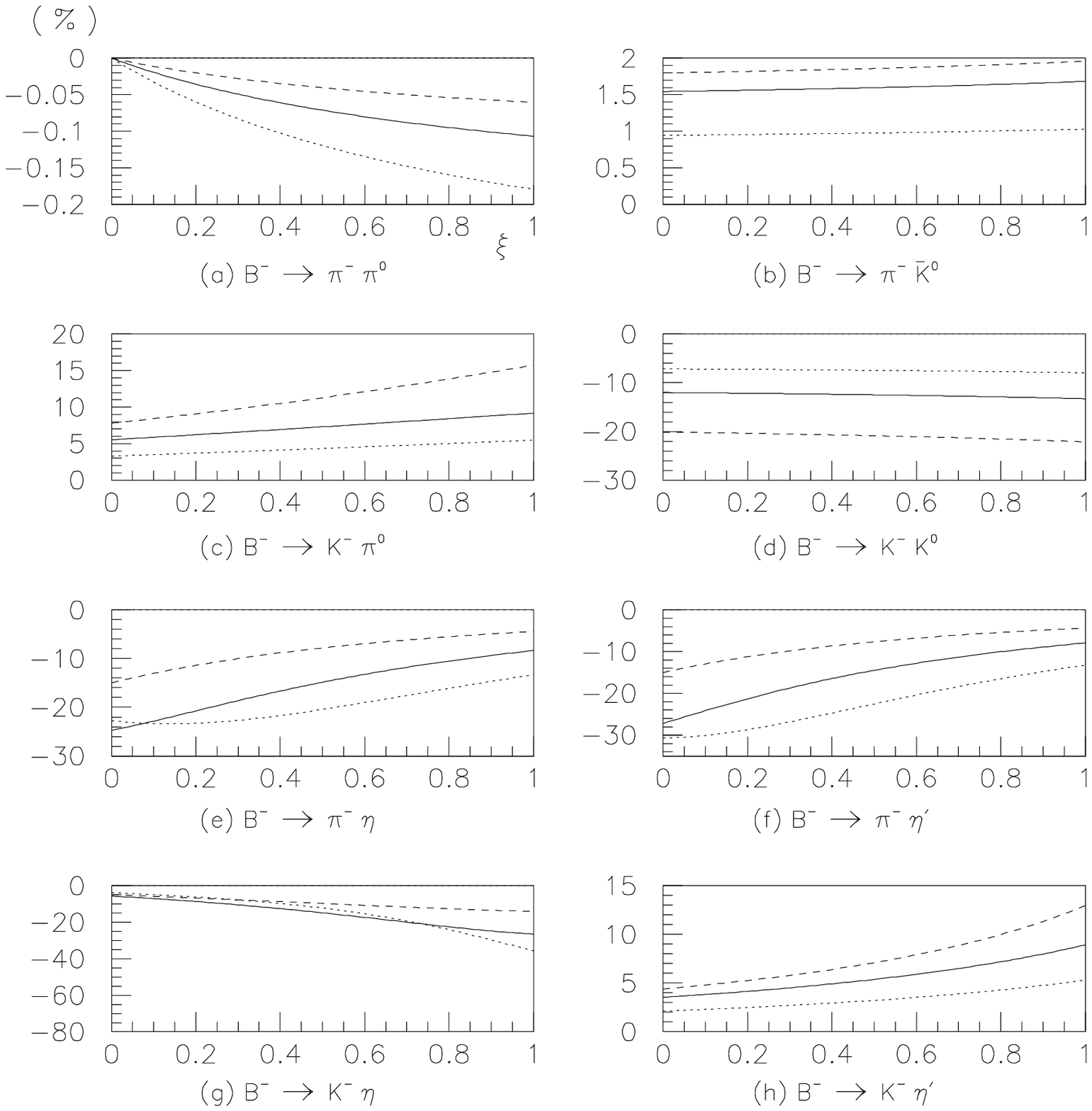}}
\caption{ The $CP$ asymmetry of the charged $B$ meson decays as
a function of nonfactorization parameter $\xi$: 
$\rho = 0.05, \eta = 0.36$ for real line, 
$\rho = 0.30, \eta = 0.42$ for dashed line and
$\rho = 0, \eta = 0.22$ for dotted line. }
\label{fig3}
\end{figure}

\begin{figure}
\centerline{\epsffile{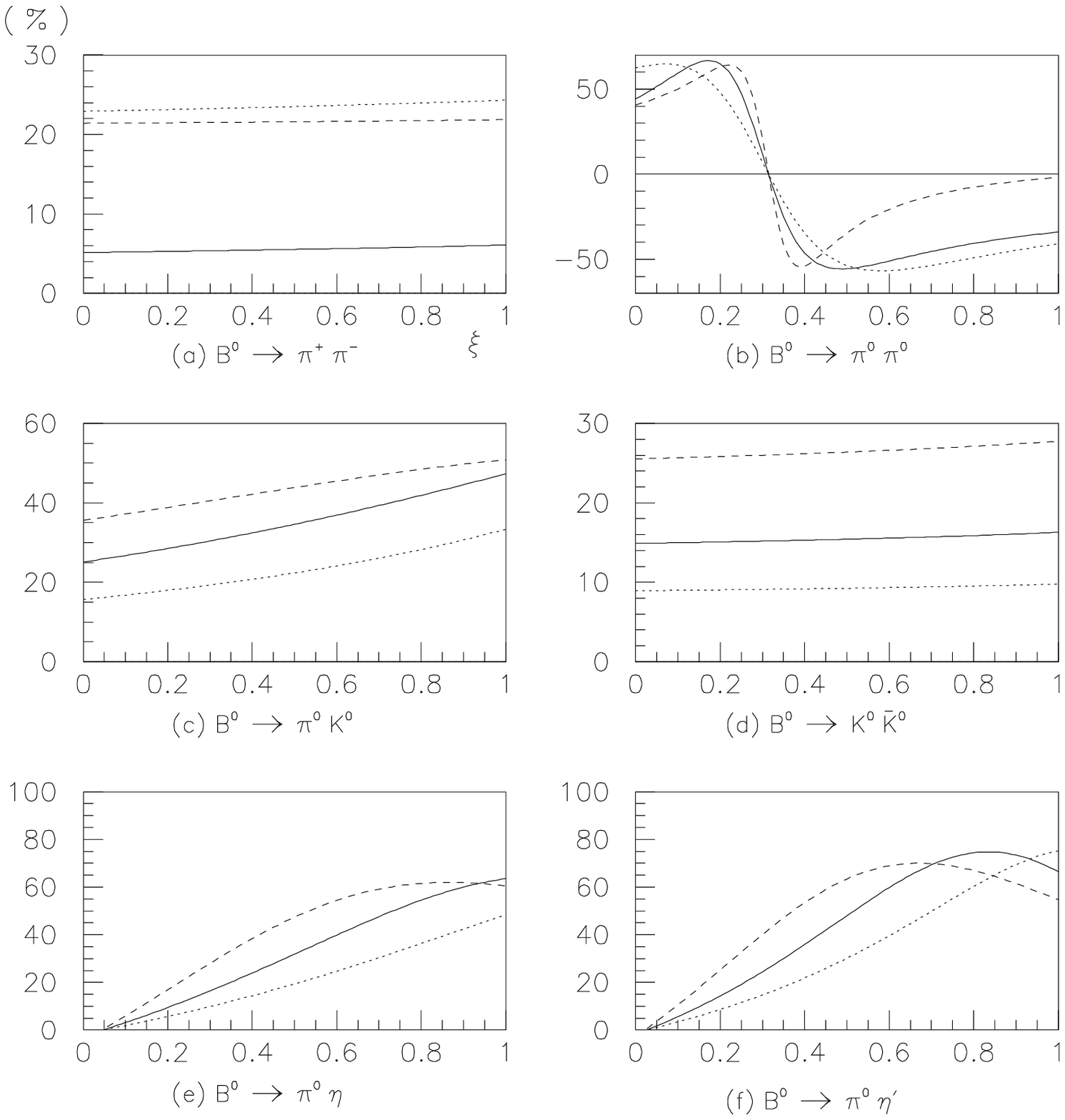}}
\caption{ The $CP$ asymmetry of the neutral $B$ meson decays as
a function of nonfactorization parameter $\xi$:  
$\rho = 0.05, \eta = 0.36$ for real line, 
$\rho = 0.30, \eta = 0.42$ for dashed line and
$\rho = 0, \eta = 0.22$ for dotted line.}
\label{fig4-1}
\end{figure}

\begin{figure}
\centerline{\epsffile{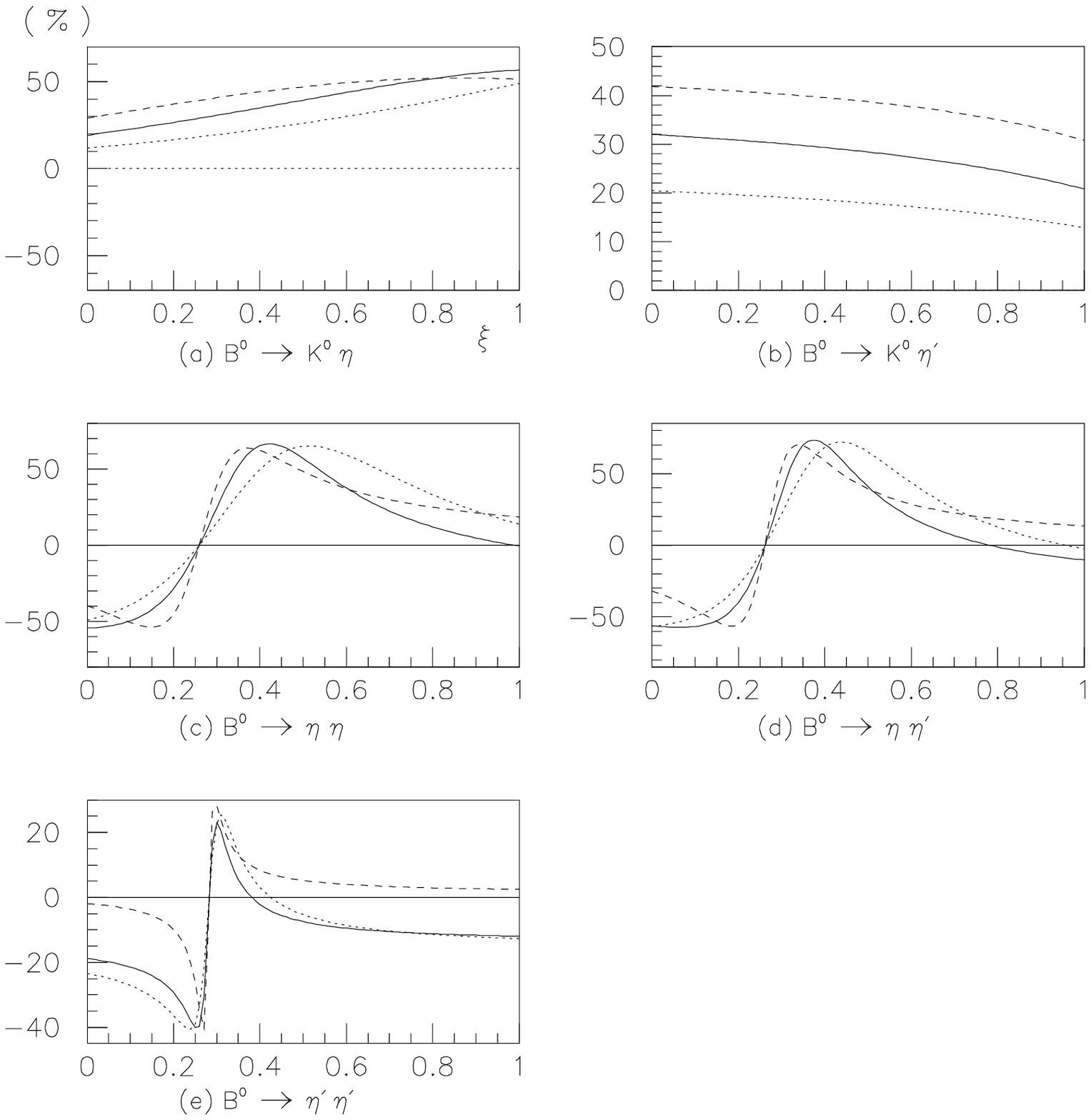}}
\caption{ The $CP$ asymmetry of the neutral $B$ meson decays as
a function of nonfactorization parameter $\xi$: 
$\rho = 0.05, \eta = 0.36$ for real line, 
$\rho = 0.30, \eta = 0.42$ for dashed line and
$\rho = 0, \eta = 0.22$ for dotted line. }
\label{fig4-2}
\end{figure}

\begin{figure}
\label{plus}
\centerline{\epsffile{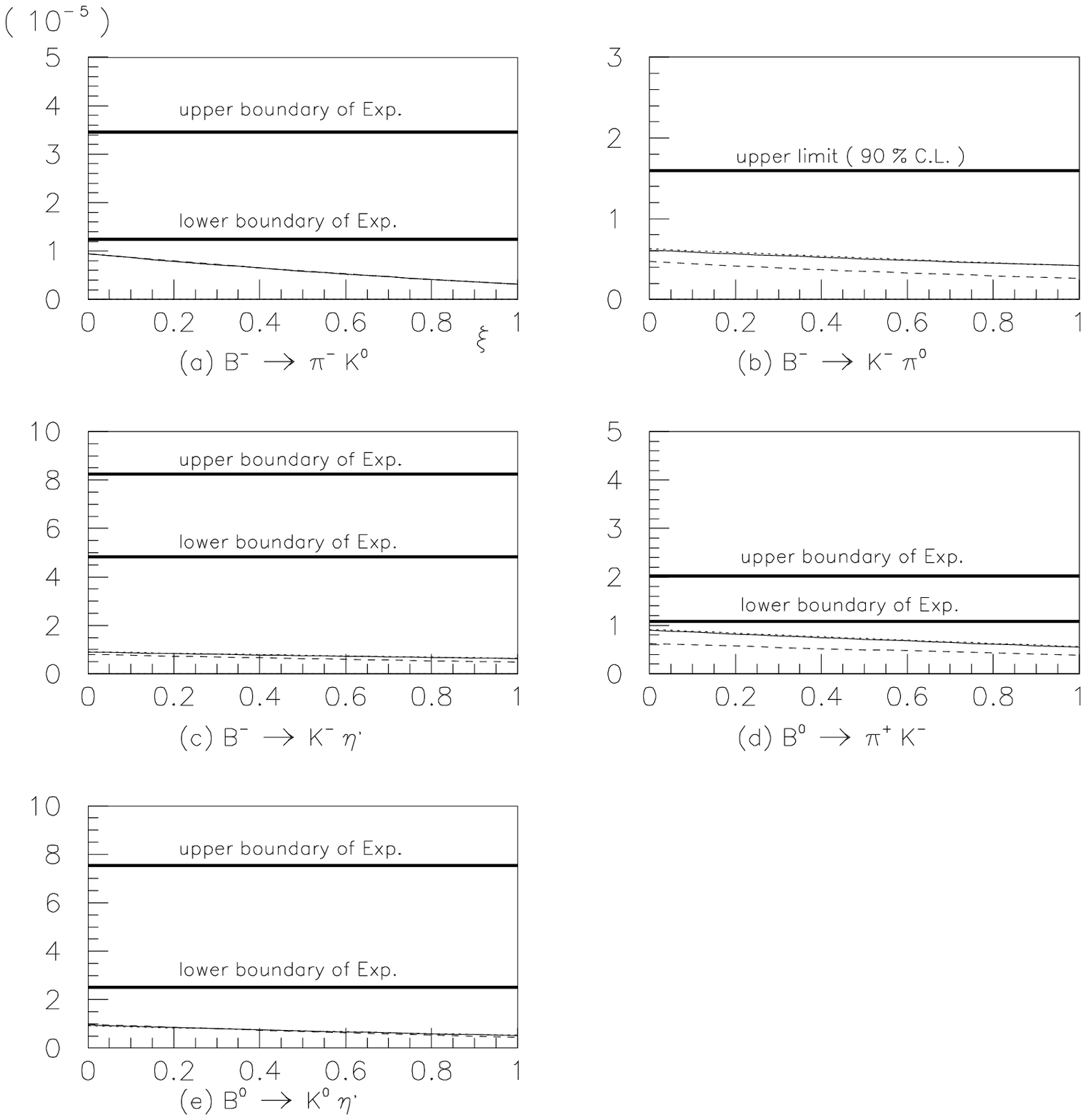}}
\caption{  
Combined branching ratios of the $B$ meson decays as functions of 
nonfactorization parameter $\xi$ in the presence of the enhanced 
$b\rightarrow s g$ with $C_{g,new} = + 5 C_{g,SM}$ at  $\mu$ = 2.5 GeV : 
$\rho = 0.05, \eta = 0.36$ for real line, 
$\rho = 0.30, \eta = 0.42$ for dashed line and
$\rho = 0, \eta = 0.22$ for dotted line. }
\label{fig5-1}
\end{figure}

\begin{figure}
\label{minus}
\centerline{\epsffile{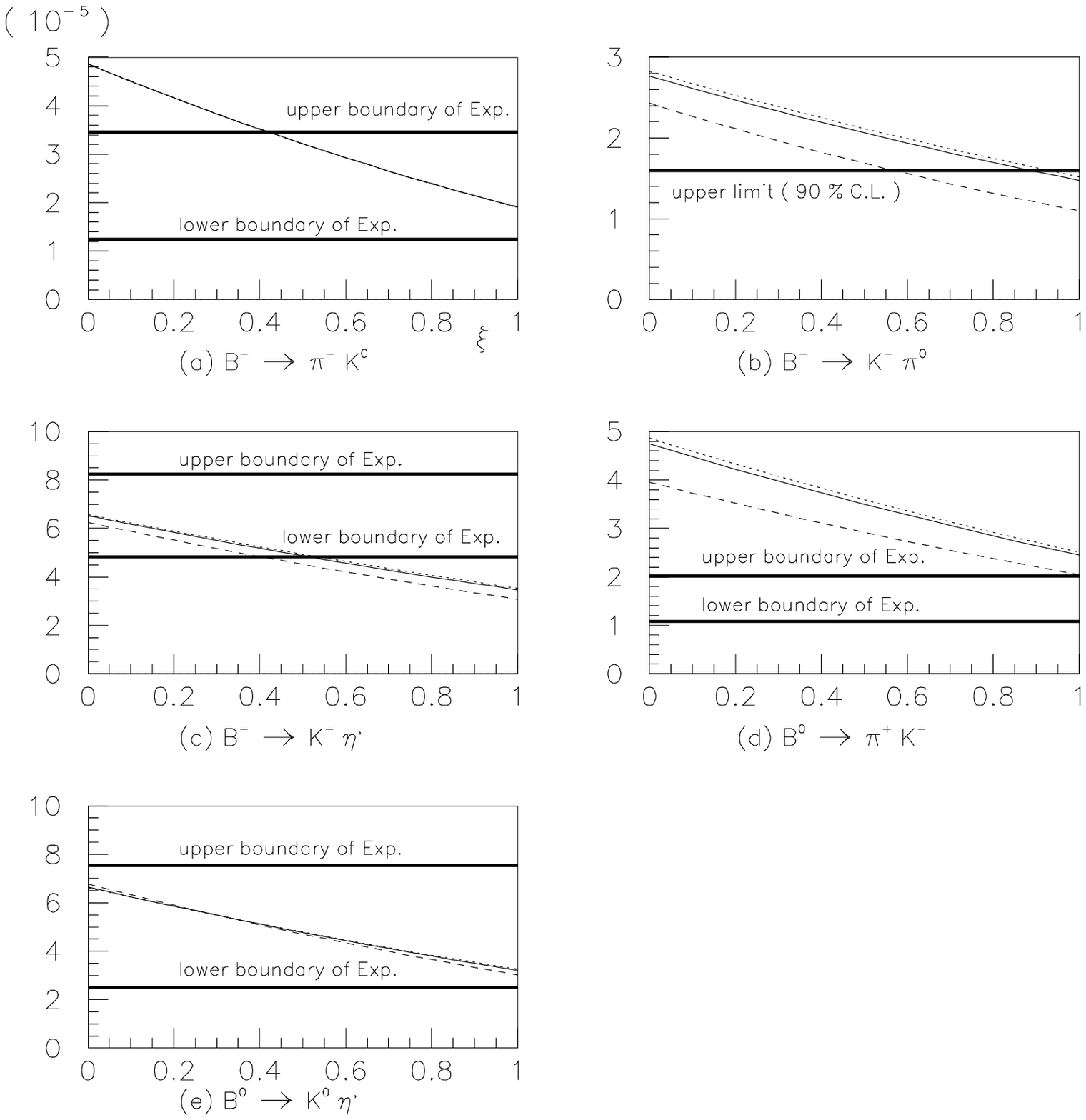}}
\caption{  
Combined branching ratios of the $B$ meson decays as functions of 
nonfactorization parameter $\xi$  in the presence of the enhanced
$b\rightarrow s g$ with $C_{g,new} = - 5 C_{g,SM}$ at  $\mu$ = 2.5 GeV : 
$\rho = 0.05, \eta = 0.36$ for real line, 
$\rho = 0.30, \eta = 0.42$ for dashed line and
$\rho = 0, \eta = 0.22$ for dotted line. }
\label{fig5-2}
\end{figure}

\begin{thebibliography}{99}

 \bibitem{CLEOpp}
R. Godang $et~ al$. (CLEO Collaboration), 
Phys. ~Rev. ~Lett. {\bf 80} (1998) 3456.

 \bibitem{CLEOeta}
B. H. Behrens $et~ al$. (CLEO Collaboration),
Phys. ~Rev. ~Lett. {\bf 80} (1998) 3710.

\bibitem{DG97}
D. Du and L. Guo,  Z.~Phys. {\bf C75} (1997) 9.

\bibitem{AG97}
A. Ali and C. Greub, Phys. ~Rev. {\bf D57} (1998) 2996.

\bibitem{ACGK97}
A. Ali, J. Chay, C. Greub and P. Ko, 
Phys. ~Lett. {\bf B424} (1998) 161.

\bibitem{DHP97}
A. Datta, X.-G. He and S. Pakvasa,
Phys. ~Lett. {\bf B419} (1998) 369.

\bibitem{CT97}
H.-Y. Cheng and B. Tseng, Phys. ~Lett. {\bf 415} (1997) 263.

\bibitem{DGR97}
A.S. Dighe, M. Gronau and J. Rosner, Phys.~Rev.~Lett. 
{\bf 79} (1997) 4333. 

\bibitem{DDO97}
N.G. Deshpande, B. Dutta and S. Oh, Phys. ~Rev. {\bf D57} (1998) 5723 ;
N.G. Deshpande, B. Dutta and S. Oh, prepint OITS-644, COLO-HEP-394, 
hep-ph/9712445.

\bibitem{ali98}
A. Ali, G. Kr\"{a}mer and Cai-Dian L\"{u}, DESY 98-041 (hep-ph/9804363)
and DESY 98-056 (hep-ph/9805403). 

\bibitem{Wil69}
K.G. Wilson, Phys.~Rev. {\bf 179} (1969) 1499.

\bibitem{CMM97}
K. Chetyrkin, M. Misiak and M. M\"unz, Phys.~Lett. 
{\bf B400} (1997) 206 ;
Erratum, ~$ibid.$ {\bf B425} (1998) 414.

\bibitem{BJLW92}
A.J. Buras $et~ al.$, Nucl. ~Phys. {\bf B370} (1992) 69.

\bibitem{neubert}
M. Neubert and B. Stech,  CERN-TH-97-099, hep-ph/9705292. 
To appear in second edition of Heavy Flavours, ed. by A.J. Buras and 
M. Lindner, World Scientific, Singapore. 

\bibitem{bjorken}
J. D. Bjorken, Nucl. ~Phys. {\bf B} ( Proc. Suppl. ) {\bf 11} (1989) 325.

\bibitem{dg91}
M. J. Dugan and B. Grinstein, Phys. ~Lett. {\bf B255} (1991) 583.

\bibitem{bpuzzle}
G. Altarelli and S. Petrarca, Phys. ~Lett. {\bf B261} (1991) 303 ;
I. Bigi $et~ al.$, Phys. ~Lett. {\bf B323} (1994) 408.


\bibitem{hou}
B.G. Grzadkowski and W.S. Hou, Phys. ~Lett. {\bf B 272} (1991) 383 ;
A. Kagan, Phys. ~Rev. {\bf D 51} (1995) 6196. ;
M. Ciuchini $et~ al.$, Phys. ~Lett. {\bf B388} (1996) 353 ;
Erratum, ~$ibid.$ {\bf B393} (1997) 489 ;
A. L. Kagan and J. Rathsman, hep-ph/9701300.

\bibitem{BSS79}
M. Bander, D. Silverman and A. Soni, Phys. ~Rev. ~Lett. {\bf 43}
( 1979 ) 242.

\bibitem{BSW87}
M. Bauer, B. Stech, M. Wirbel, Z. ~Phys. {\bf C34} ( 1987 ) 103. 

\bibitem{Leut97}
H. Leutwyler, Nucl. ~Phys. ~Proc. ~Suppl. {\bf 64} (1998) 223.

\bibitem{FK97}
T. Feldmann and P. Kroll, preprint WUB 97-28, hep-ph/9711231 ;
T. Feldmann, P. Kroll and B. Stech, WU-B-98-2, hep-ph/9802409.

\bibitem{Wol83}
L. Wolfenstein, Phys. ~Rev. ~Lett. {\bf 51} ( 1983 ) 1945.


\bibitem{Ali96}
A. Ali, Acta ~Phys. ~Polon. {\bf B27} ( 1996 ) 3529.


\bibitem{HZ97}
I. Halperin and A. Zhitnitsky, 
Phys. Rev. {\bf D56} (1997) 7247 ;
Phys. Rev. Lett. {\bf 80} (1998) 438 ;
hep-ph/9706251.

\bibitem{Wol91}
L. Wolfenstein, Phys. ~Rev. {\bf D43} ( 1991 ) 151.

\bibitem{GH91}
J. Gerard and W. Hou, ~Phys. ~Rev. {\bf D43} ( 1991 ) 2909.

\bibitem{KPS95}
G. Kramer, W. F. Palmer and H. Simma, Z. ~Phys. {\bf C66}
( 1995 ) 429.

\bibitem{BFM97}
A. Buras, R. Fleischer and T. Mannel, CERN-TH/97-307, TUM-HEP-300/97,
TTP97-41, hep-ph/9711262.

\bibitem{Neu97}
M. Neubert, Phys. ~Lett. {\bf B424} (1998) 152 ;
D. Atwood and A. Soni, hep-ph/9712287 (1997) ;
A. F. Falk, A. L. Kagan, Y. Nir and A. A. Petrov,
Phys. Rev. {\bf D57} (1998) 4290.  
J.M. Gerard and J. Weyers,  UCL-IPT-97-18, hep-ph/9711469 ;
D. Delepine, J.M. Gerard, J. Pestieau and J. Weyers, UCL-IPT-98-01, 
hep-ph/9802361. 

\bibitem{Flei98}
R. Fleischer, CERN-TH/98-60, hep-ph/9802433.

\bibitem{FM98}
R. Fleischer and T. Mannel, Phys. ~Rev. {\bf D57} ( 1998 ) 2752.

\bibitem{Gro89}
M. Gronau, Phys. ~Rev. ~Lett. {\bf 63} ( 1989 ) 1451.

\bibitem{BHP96}
T. E. Browder $et~ al.$, hep-ph/9606354 ;
T. E. Browder, hep-ph/9611373 ;
J. D. Richman, hep-ex/9701014.

\bibitem{Barish96}
B. Barish $et~ al.$, Phys. ~Rev. ~Lett. {\bf 76} (1996) 1570 ;
H. Albrecht $et~ al.$, Phys. ~Lett. {\bf B318} (1993) 397.

\bibitem{BBBG94}
E. Bagan $et~ al.$, Nucl. ~Phys. {\bf B432} (1994) 3 ;
Phys. ~Lett. {\bf B342} (1995) 362 ; Erratum, ~$ibid.$ 
{\bf 374} (1996) 363. 

\bibitem{Browder98}
T. E. Browder $et~ al.$, hep-ex/9804018.

\bibitem{LNO97}
A. Lenz $et~ al.$, Phys. ~Rev. {\bf D56} (1997) 7228.
\end{thebibliography}
\end{document}